\documentclass[epj]{svjour}
\usepackage{graphicx}
\usepackage[usenames,dvipsnames]{color}
\usepackage{units}
\usepackage{subfigure}
\usepackage[pdfpagelabels,bookmarks=true,colorlinks,linkcolor=black,urlcolor=black,citecolor=blue]{hyperref}
\usepackage{memhfixc}

\newcommand{\KE}{KATRIN experiment}
\newcommand{\KM}{KATRIN main spectrometer}

\newcommand{\Mac}{MAC-E filter}

\newcommand{\ie}{\emph{i.e.}}
\newcommand{\eg}{\emph{e.g.}}
\newcommand{\belec}{$\beta$ electron}
\newcommand{\cps}{counts/s}

\begin{document}

\title{Effect of a sweeping conductive wire on electrons stored in a Penning-like trap between the KATRIN spectrometers}

\author{M. Beck\inst{1} \and K. Valerius\inst{1}\thanks{Present address: Physikalisches Institut, Friedrich-Alexander-Universit\"at Erlangen-N\"urnberg, Germany} \and J. Bonn\inst{2} \and K. Essig\inst{3} \and F. Gl\"uck\inst{4,5} \and H.-W. Ortjohann\inst{1} \and B. Ostrick\inst{1,2} \and E. W. Otten\inst{2} \and Th. Th\"ummler\inst{3}\thanks{Present address: Institut f\"ur Kernphysik, Karlsruhe Institute of Technology, Germany} \and M. Zbo\v{r}il\inst{1,6} \and C. Weinheimer\inst{1,3}}

\institute{Institut f\"ur Kernphysik, Westf\"alische Wilhelms-Universit\"at M\"unster, Germany \and
Institut f\"ur Physik, Johannes Gutenberg-Universit\"at Mainz, Germany \and
Helmholtz-Institut f\"ur Strahlen- und Kernphysik, Rheinische Friedrich-Wilhelms-Universit\"at Bonn, Germany \and 
Institut f\"ur Experimentelle Kernphysik, Karlsruhe Institute of Technology, Germany \and 
Res. Inst. Nucl. Part. Phys., Budapest, Hungary \and 
Nuclear Physics Institute ASCR, \v{R}e\v{z} near Prague, Czech Republic}

\mail{marcusb@uni-muenster.de}

\abstract{
The KATRIN experiment is going to search for the mass of the electron
antineutrino down to $\unit[0.2]{eV/c^2}$. In order to reach this sensitivity the
background rate has to be understood and minimised to $0.01$~counts per
  second. One of the background sources is the unavoidable Penning-like trap for electrons
due to the combination of the electric and magnetic fields
between the pre- and the main spectrometer at KATRIN. 
In this article we will show that by sweeping a conducting wire
periodically through such a particle trap stored particles can be removed, 
an ongoing discharge in the trap can be stopped, and the count rate measured with a detector looking at the trap is reduced.
}

\titlerunning{Effect of a sweeping wire}
\authorrunning{M. Beck et al.}

\maketitle

\section{Motivation}
\label{sec:motivation}

\begin{figure*}[!htb]
 \centering
\includegraphics[width=\textwidth]{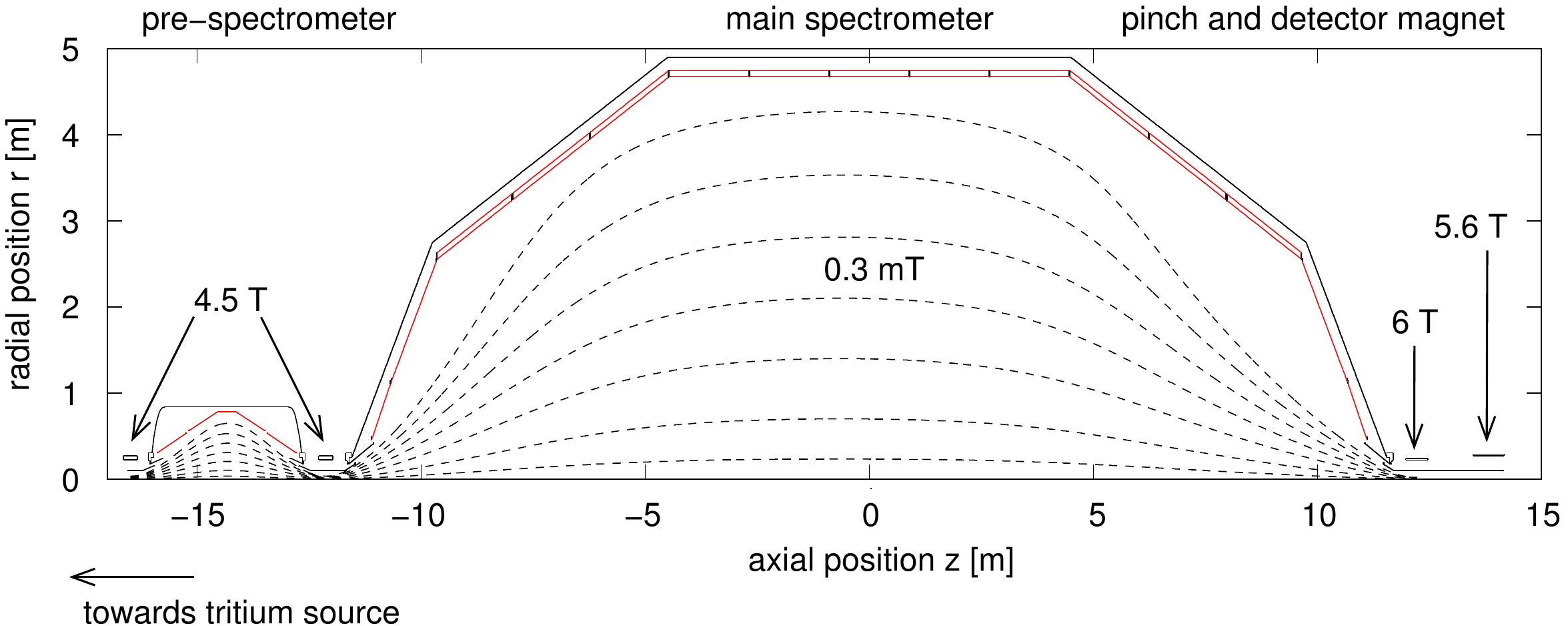}
\caption[Overview of high-field regions in the KATRIN spectrometer and
detector section]{Overview of the two-spectrometer set-up of the KATRIN experiment (left: small 
pre-spectrometer, right: large main spectrometer) and its magnetic fields according to the present KATRIN default configuration. Note that in the original KATRIN design \cite{kdr} two short solenoids
and a 3~m long magnetic transport section were foreseen between the two
spectrometers. These two transport magnets are replaced by a single short solenoid in the present KATRIN default configuration. Dashed lines represent magnetic field lines. The field lines in the main spectrometer are shaped by an air coil system which is not shown. High voltage of the order of -18.6~kV is applied in the centre (weak magnetic field) regions of the spectrometers (pre-spectrometer: $z \approx -15$~m, main spectrometer: $z \approx 0$~m). At the magnet between the spectrometers at $z=-12$~m the
  electric potential is zero. In combination with the magnetic field this potential well creates a large Penning-like trap for electrons between the pre- and the main spectrometer. \label{fig:lineup-trapping}}
\end{figure*}

In light of the confirmation  of neutrino oscillations by many experiments in the last decade, 
the question of the absolute mass scale of neutrinos is very important for particle
physics and cosmology (see \eg, \cite{otten-weinh-review,lesgourgues-pastor-review}).
The KArlsruhe TRItium Neutrino experiment
(KATRIN, \cite{kdr}) aims to search for the mass of the electron antineutrino with
a sensitivity of $\unit[0.2]{eV/c^2}$. At KATRIN electrostatic energy filters of
{\Mac} type (electrostatic filter with magnetic adiabatic collimation, \cite{picard-nimb,lobashev85}) are used both for the precision measurement of the endpoint of
the beta decay of tritium with the main spectrometer, and for the reduction
of the bulk of the electrons at energies of approx. $\unit[200]{eV}$
below the endpoint with the pre-spectrometer, which is located upstream of the
main spectrometer (see fig.~\ref{fig:lineup-trapping}).

A spectrometer of {\Mac} type  consists of an electrostatic retardation potential, acting as a high pass filter, combined with an inhomogeneous magnetic guiding field. Two superconducting solenoids produce a strong magnetic field $B_\mathrm{max}$ at the entrance and exit of the spectrometer. The magnetic field strength decreases towards the centre of the spectrometer to a minimum value\footnote{In the KATRIN set-up, this reduction factor $B_\mathrm{max}/B_\mathrm{min}$ amounts to 
20000.} $B_\mathrm{min}$. This results in an expansion of the magnetic
flux tube towards the centre of the spectrometer. Due to the magnetic gradient
force the radial energy $E_\perp$ of an
electron that moves adiabatically into this weak field area is converted nearly completely into longitudinal energy $E_\parallel$
according to the conservation of the magnetic orbital momentum 
\begin{equation}
  \mu = \frac{E_\perp}{B}
\end{equation}
(equation given in the non-relativistic limit). The longitudinal energy is then probed by the electric potential $U_\mathrm{spec}$ in the analysis plane at $B = B_{\mathrm{min}}$ and $z = 0$ (see fig.~\ref{fig:lineup-trapping}).
Only electrons with sufficient longitudinal
kinetic energy, $E_\parallel > qU_\mathrm{spec}$, with $q=-e$ being the
electron charge, will be able to pass the filter and get re-accelerated
towards the detector at the exit of the spectrometer.

To reach the proposed sensitivity at KATRIN the background rate at the endpoint of the tritium beta decay has to be minimised. One background component at KATRIN
stems from particle traps that  exist due to the
combination of the high electric and magnetic fields of the {\Mac}s themselves. Through ionisation processes a trapped particle can cause an increase of the number of charged particles, 
which in turn may give rise to an enhanced background count rate.

\begin{figure}
 \centering
\includegraphics[width=0.4\textwidth]{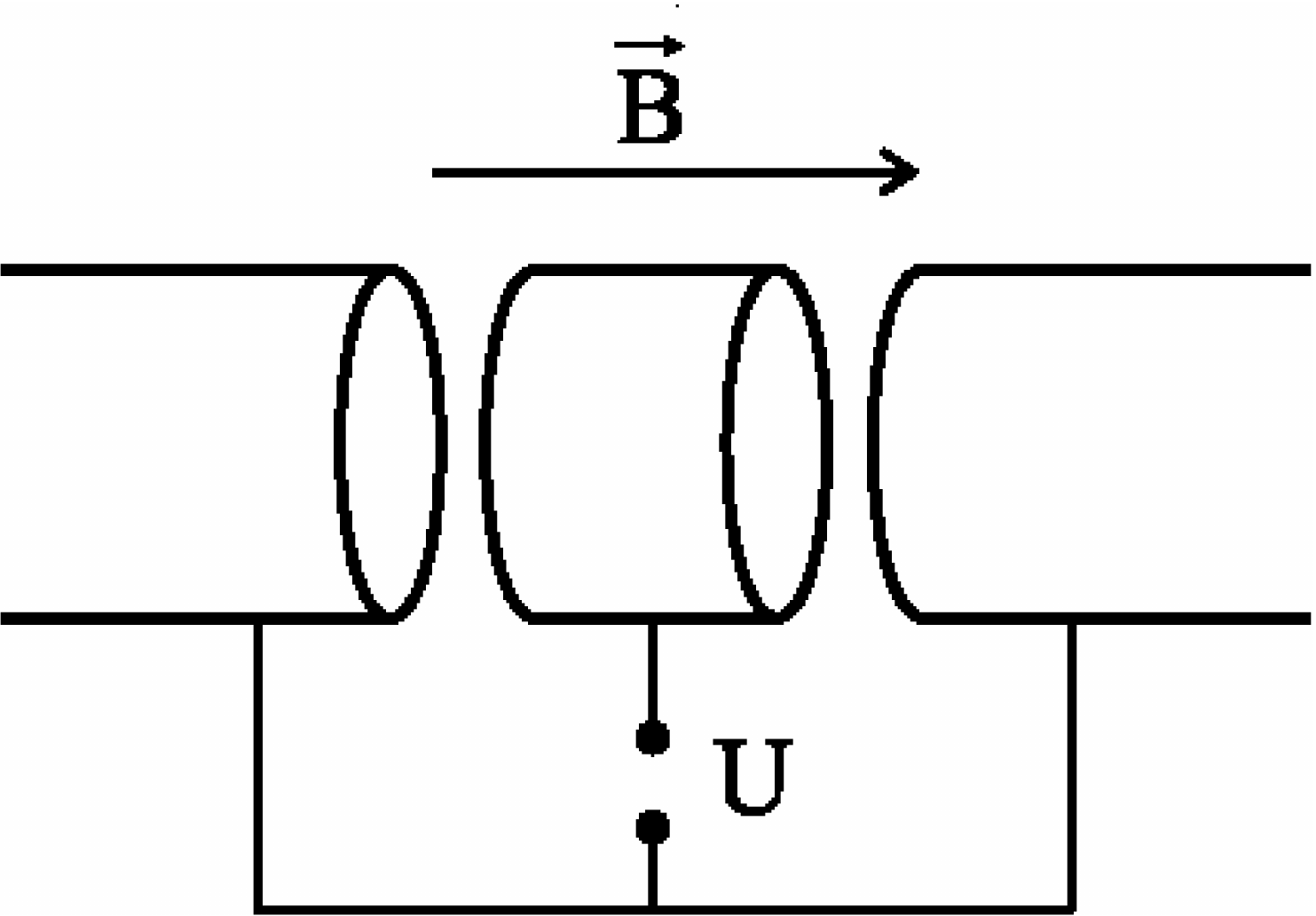}
\caption[Principle of a cylindrical Penning trap]{Principle of a cylindrical
  Penning trap. The electrons are confined by an electrostatic potential U in
  axial direction and by a B-field in radial direction.  \label{fig:penning}}
\end{figure}

One type of particle trap of particular importance for  KATRIN is the Penning trap \cite{blaum_review}. In a Penning trap charged particles are confined by a magnetic field in radial direction. The confinement in axial direction is achieved by an electrostatic potential (see fig.~\ref{fig:penning}).
The {\Mac} for electrons with its negative central potential and both sides at
ground potential together with the axial magnetic field constitutes a large Penning-like trap for positively charged particles. Light positively charged particles do not exist in the KATRIN experiment\footnote{There is no known source of positrons at KATRIN.} and for more massive positively charged particles, \eg, protons, the cyclotron radii typically get too large so that the charged particles can escape radially.
 
Penning-like traps for electrons, however, can easily be created at various locations inside the \Mac. Great care has to be taken in the design of the electric and magnetic fields in order to avoid these traps \cite{diss_kathrin,diss_habermehl}. 
In a two-spectrometer set-up like the one formed by the KATRIN pre- and main
spectrometer, such a trap for electrons cannot be avoided in principle.
The two negative retarding potentials together with the high magnetic field between them form a large Penning-like trap for electrons (compare fig.~\ref{fig:lineup-trapping}).

In this paper we will focus on the particular kind of electron trap caused by the combination of the two spectrometers of \Mac\ type. 
We will report in section \ref{sec:background} on our first estimates whether and how this trap could cause a significant background rate 
at the KATRIN experiment. 
In sections \ref{sec:experimental_setup} and \ref{sec:results} we will present our experimental set-up and the investigations of our new method to periodically empty this electron trap in order to reduce the background rate which may originate from electrons stored in this trap. In section \ref{sec:discussion} we will give our conclusions. 

\section{Background from the inter-spectrometer trap}
\label{sec:background}

We expect that the electron trap, which is formed by the combination of the
pre- and the main spectrometer of KATRIN, will be filled from the outside
at a rate comparable to the total background rate on the detector, which is expected to be $\approx 0.2$~\cps \footnote{At the Mainz neutrino mass experiment the background rate in the energy window of interest was of order 0.01~\cps, whereas the background over the full energy range amounted to about 0.1~\cps. Although the KATRIN pre- and main spectrometer are larger and hence should exhibit a larger background, we expect that the advanced electrostatic shielding by a wire electrode system \cite{kdr,diss_kathrin} will allow to keep the spectrometer-related background rate at the same level as in the Mainz experiment. For the considerations presented in this paper we thus assume that the two KATRIN spectrometers will emit electrons towards both of their ends at a rate comparable to the 0.1~\cps\ determined at Mainz.}.
These electrons confined within the inter-spectrometer trap are not a direct source of background, but can produce indirect background at the KATRIN detector.
Three known processes could create such background. Two of them are connected to ionisation processes in the trap and positively charged ions leaving the trap
towards the main spectrometer, where they produce secondary electrons in the volume or at the walls. 
The third process starts with a photon from deexcitation or recombination processes in the trap, which creates also secondary electrons in the main spectrometer. 
All three processes may yield similar background rates and studies of these effects are still in progress. 
In the following we will elaborate on the first process, background electrons from ionisation of rest gas due to ions from the trap. This is a two-step process:

\begin{enumerate}
\item An electron stored in the trap interacts with the residual gas, creating a secondary electron and a positively charged ion. In most cases the former will get trapped, as well. The latter, which usually will be a proton (H$^+$) or a H$_2^+$, will leave the trap and will be accelerated by the electric field towards either the pre- or the main spectrometer. 
\item In the low magnetic field $B_\mathrm{min} = 0.3$~mT in the centre of the main spectrometer the cyclotron radius of the ion will become too big and hence the ion will not be guided adiabatically by the magnetic field. Thus it will be lost by hitting a spectrometer wall\footnote{When hitting the wall the ion could eject a secondary electron, which has a very small chance to reach the detector. This is the second background processs mentioned above. The natural magnetic shielding of a MAC-E filter together with the electric shielding by a two-layer wire electrode system \cite{kdr,diss_kathrin} suppress this process by about 7 orders of magnitude.}, as confirmed by our particle tracking simulations\footnote{Ions from the trap that reach the pre-spectrometer will not be trapped there either, since they will have gained sufficient energy in the electric potential to move non-adiabatically and will therefore hit the pre-spectrometer electrodes or wall and thus will be removed from the trap.}.
On its about 20~m long way until hitting the wall it has only a minor chance
to cause a second ionisation. But if this happens in the high retarding
potential of the main spectrometer, there is a 50~\% chance that the created
tertiary electron will follow the gradient of the electric field towards the
detector exit of the main spectrometer and be counted as a background
electron. Due to the limited energy resolution of the electron detector
($\Delta E_\mathrm{det} \approx 1$~keV) this electron cannot be distinguished
from a signal electron, because it will reach the detector with a kinetic
energy given by the sum of its energy at creation $E_\mathrm{start} =
{\cal{O}}(10)$~eV and the retarding potential $qU = 18.6$~keV. Since the
signal electrons from the endpoint region of tritium beta decay with an energy
a little bit above the retarding potential $qU$ have about the same energy
they cannot be distinguished from these background electrons.

The cross section for the processes 
\begin{equation}
  \mathrm{H}^+ + \mathrm{H}_2 \rightarrow \mathrm{e^-} + X,\, \qquad 
  \mathrm{H}_2^+ + \mathrm{H}_2 \rightarrow \mathrm{e^-} + X\, \qquad
\end{equation}
amounts to $\sigma_\mathrm{ion} \approx 10^{-16}$~cm$^2$ at $E_\mathrm{ion} \approx 18.6$~keV \cite{and00}. This cross section results in an ionisation probability $P_\mathrm{ion}$ for an ion track length of $l=20$~m and a residual gas pressure\footnote{At the ultra-high vacuum level of the KATRIN experiment the residual gas mainly consists of H$_2$ molecules.} of $p = 10^{-11}$~mbar ($n(\mathrm{H}_2) = 2.5\cdot 10^{11}$/m$^3$) of
\begin{equation}
  P_\mathrm{ion} = \sigma_\mathrm{ion} \cdot n \cdot l = \frac{\sigma_\mathrm{ion} \cdot p \cdot l}{k_\mathrm{B} \cdot T} = 5 \cdot 10^{-8}
\end{equation} 
\end{enumerate}

This probability is very low and one might tend to neglect the described chain of processes in case there is no other multiplication process.
However, a secondary electron created by the first ionisation in the trap has a significant chance of being created in a high potential at either end of the trap, the reason being that the ionisation cross section is highest when the primary electron is low in energy. This condition is fulfilled close to the reflecting retardation potential. In this case the secondary electron gains enough energy in the electric potential to perform another ionisation itself. Therefore, we have to check how many ionisations $N_\mathrm{ion}$ in the trap can be traced back to a single trapped electron. If $N_\mathrm{ion}$ is of order $1/P_\mathrm{ion}$ this chain of background processes could lead to a significant background rate at the detector. 

In the following, we report on first trajectory simulations of electrons in this Penning-like inter-spectrometer trap including cooling processes by synchrotron radiation as well as by elastic and inelastic scattering of electrons off the residual gas \cite{dipl_essig}.

\begin{figure}[tb]
 \centering
 \includegraphics[width=0.485\textwidth]{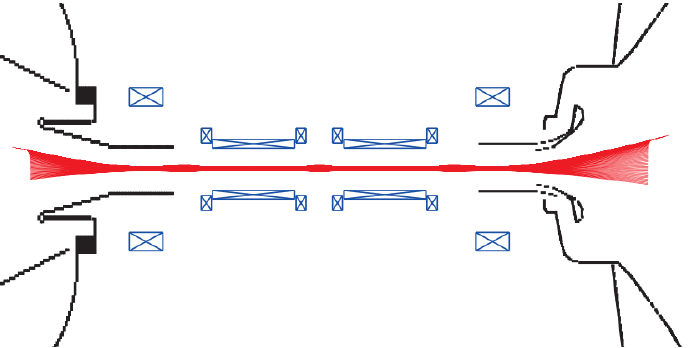}
  \caption{Example of a tracking simulation of a trapped electron in the
    inter-spectrometer electron trap \cite{dipl_essig}. The experimental
    set-up corresponds to the description in the KATRIN design report \cite{kdr}
    with two $2~\mathrm{m}$ long transport magnets ($B = 5.4~\mathrm{T}$) and two short solenoids.}
  \label{fig:tracking}
\end{figure}

The electron tracking calculations (an example is shown in fig. \ref{fig:tracking}) were performed with the program package Adipark \cite{thuemmler_dipl} using the guiding-centre approximation:
 In zeroth approximation the \belec s are spiralling around the guiding magnetic field lines. Therefore, we used the magnetic field lines calculated by the program Bfield3d \cite{flatt_dipl} to propagate the electrons, while adjusting the energy according to the local electrostatic field as computed by the program SimIon 7.0 \cite{simion}. Additionally, in non-homogeneous electric and magnetic fields the electrons feel a small drift $u$, which to first order \cite{picard-nimb} ($c=1$) reads
\begin{eqnarray}
  \vec u = \left( \frac{\vec E \times \vec B}{B^2} -\frac{(E_\perp + 2 E_{||})}{e\cdot B^3}
(\vec B \times \nabla_\perp \vec B) \right). 
\label{eq:drift}
\end{eqnarray}
It results in a motion transverse to the above-mentioned component along the magnetic field line. In each tracking step the energy loss by synchrotron radiation was taken into account by using the continuous synchrotron radiation formula \cite{jackson}
\begin{equation}
  \frac{1}{\tau_\mathrm{Sy}} = \frac{\dot E_\perp}{E_\perp} = 0.4~ {\rm s}^{-1} \cdot
              \left( \frac{B}{\rm 1~T} \right)^2.
  \label{eq:synchrotron}
\end{equation}

\begin{figure}[tb]
 \centering
 \includegraphics[width=0.485\textwidth]{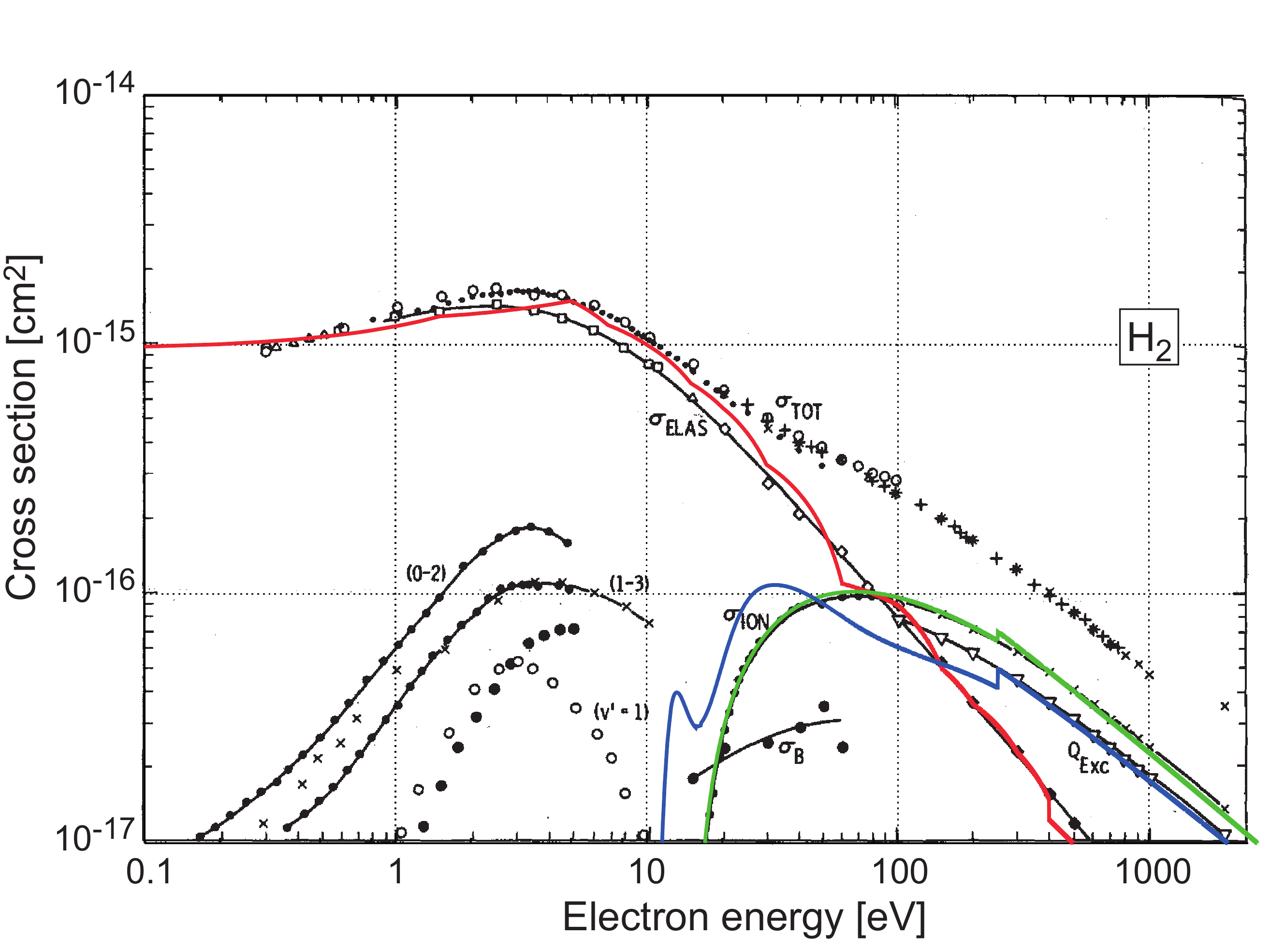}
  \caption{Comparison of the total cross sections used in our simulation with
  experimental data \cite{tra83}. The total cross sections for elastic processes of electrons on H$_2$ molecules
  are shown in red,
  for molecular excitation in blue and for ionisation in green \cite{dipl_essig}.}
  \label{fig:scattering_xsection}
\end{figure}

\begin{figure}[tb]
 \centering
 \includegraphics[width=0.485\textwidth]{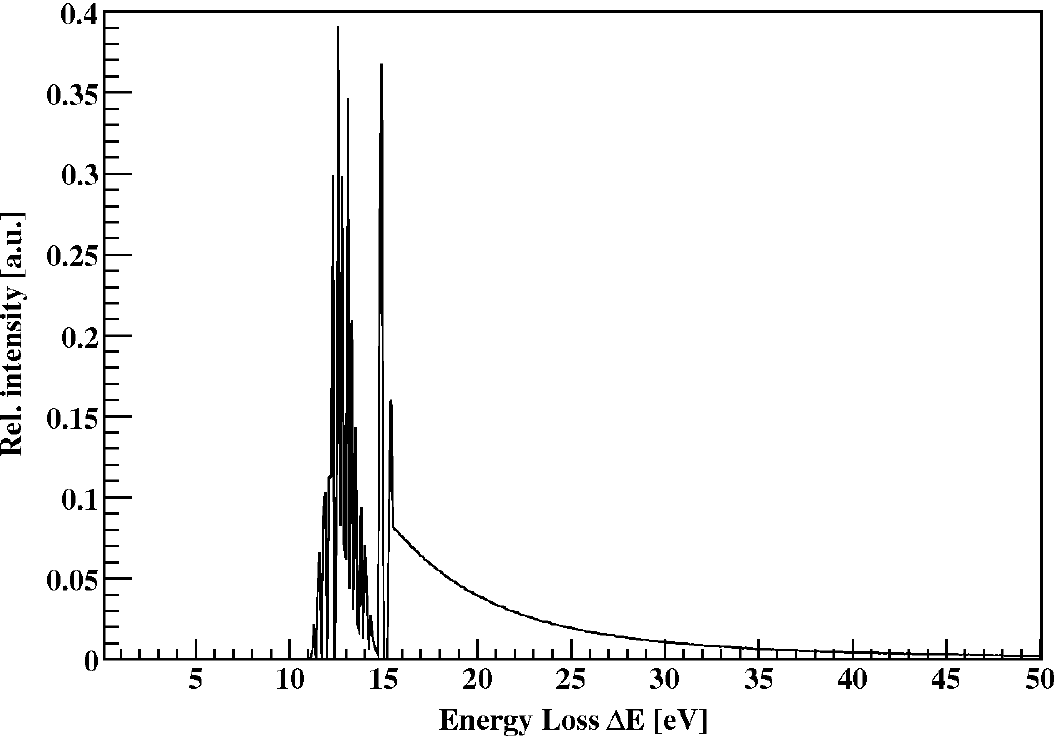}
  \caption{Energy loss $\Delta E$ for molecular excitations and ionisations of 18~keV electrons on H$_2$ averaged over all scattering angles $\vartheta$ according to our energy loss model \cite{dipl_wolff,glueck10}.}
  \label{fig:energy_losses}
\end{figure}

The elastic and inelastic scattering on the H$_2$ molecules of the residual gas is simulated
as a random process using the total cross sections from refs.~\cite{tra83,li87} for elastic scattering,
\cite{arr80,zhi95} for molecular excitation, and  \cite{cp96,pr73} for ionisations. Figure 
\ref{fig:scattering_xsection} compares these cross sections with experimental data.
Each scattering process is related to an energy loss $\Delta E$ and a scattering
angle $\vartheta$, which can be described by a detailed energy loss model (see fig.~\ref{fig:energy_losses}) to be discussed in a forthcoming publication \cite{glueck10}. As the transverse energy rapidly decreases due to synchrotron radiation, special care has been taken with regard to the realistic description  of the scattering angle. For elastic scattering the correlation
between energy loss $\Delta E$ and scattering angle $\vartheta$ reads
\begin{equation}
  \Delta E = 2 \cdot \frac{E_{\mathrm{kin,e}} \cdot m_{\mathrm e}}{m_{\mathrm H_2}} \cdot \left( 1- \cos{\vartheta} \right).
\label{eq:scatangle}
\end{equation}
For large energy losses $\Delta E > 100$~eV by ionisation the out-going electron can be treated as quasi-free before the scattering process. Therefore, the correlation 
between energy loss $\Delta E$ and scattering angle $\vartheta$ is given by
\begin{equation}
  \Delta E = E_{\mathrm{kin,e}} \cdot \left( 1- \cos^2{\vartheta} \right).
\end{equation}
For smaller energy losses this approximation is no longer valid and we refer
to the description of our detailed energy loss model, which will be discussed
in \cite{glueck10}. We would like to note that our energy loss model agrees
with exerimental data on the energy loss of $18~\mathrm{keV}$  electrons on
molecular hydrogen \cite{troitsk00}.

\begin{figure}[tb]
 \centering
 \includegraphics[width=0.485\textwidth]{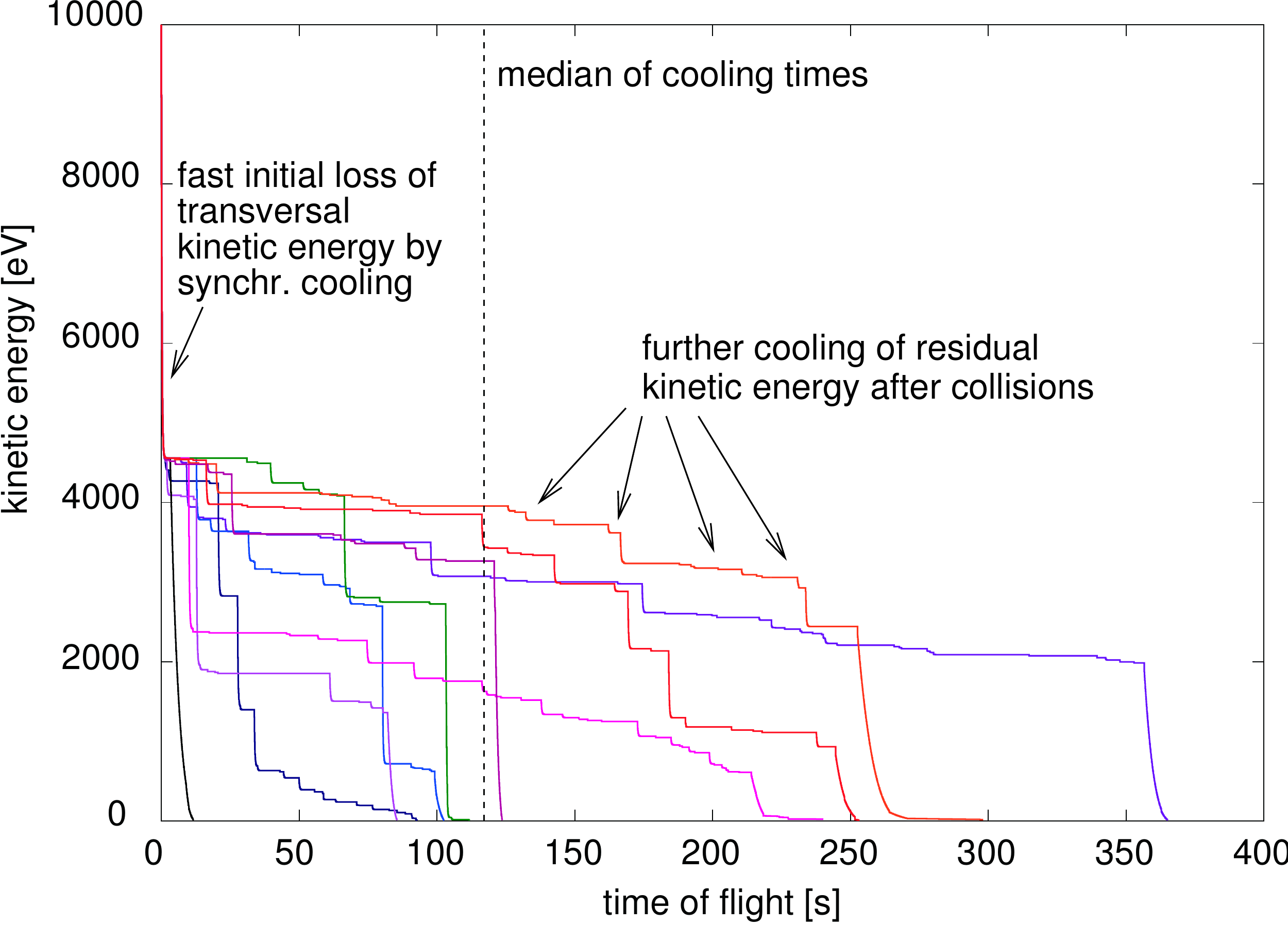}
  \caption{Simulated energy loss of 10 electrons within
  the Penning-like trap between the KATRIN pre- and main spectrometers \cite{dipl_essig}.
  Energy losses by synchrotron radiation as well as by elastic and
  inelastic scattering processes on residual H$_2$ molecules
  were considered. The electrons
  started with an energy of $E(0)= 10$\,keV halfway between the two
  spectrometers at a radius of $r=0.5$\,cm and an angle of
  $\theta=45^\circ$ with respect to the magnetic field lines.
  The simulations were performed for a residual gas pressure of
  $p($H$_2) = 10^{-10}$\,mbar, \ie, 10 times higher than expected for KATRIN.
  After each collision event the transverse energy is lost by synchrotron
  radiation within a few seconds, in addition to the energy transfer to the collision partner
 (sharp steps in the curves). It takes about 100\,s (the median of the 10 electrons is 120~s)
  to cool the electrons down to the ionisation threshold of 15.4~eV.}
  \label{fig:ionization_cooling}
\end{figure}

Figure~\ref{fig:ionization_cooling} presents the energy loss obtained for the simulation of a sample of 10 trapped electrons in the old KATRIN design (see fig.~\ref{fig:tracking}). The simulation results show that
these electrons lose their transverse energy after each collision
event by synchrotron radiation according to eq.~(\ref{eq:synchrotron}) 
on a time scale of less than a second due to the high magnetic field of 5.4\,T in the transport magnets. 

At a rest gas pressure of $10^{-10}$~mbar (for technical
reasons chosen 10 times higher than that expected for KATRIN), the scattering
processes on the rest gas occur at a rate of about 1 per 4\,s per
stored electron (see fig.~\ref{fig:ionization_cooling}). As indicated in 
fig.~\ref{fig:scattering_xsection} most
processes are elastic or excitation processes without ionisation.
Like synchrotron radiation, elastic scattering and excitation
processes cool the trapped electron without creating dangerous secondary
electron--ion pairs. It should be noted that elastic scattering
plays an important role in this cooling process although its corresponding energy loss is small. However, the scattering
angle might be large (see eq.~\ref{eq:scatangle}), redistributing energy from
longitudinal to transverse motion, which is radiated away quickly.

\begin{figure}[htb]
\centering
 \includegraphics[width=0.485\textwidth]{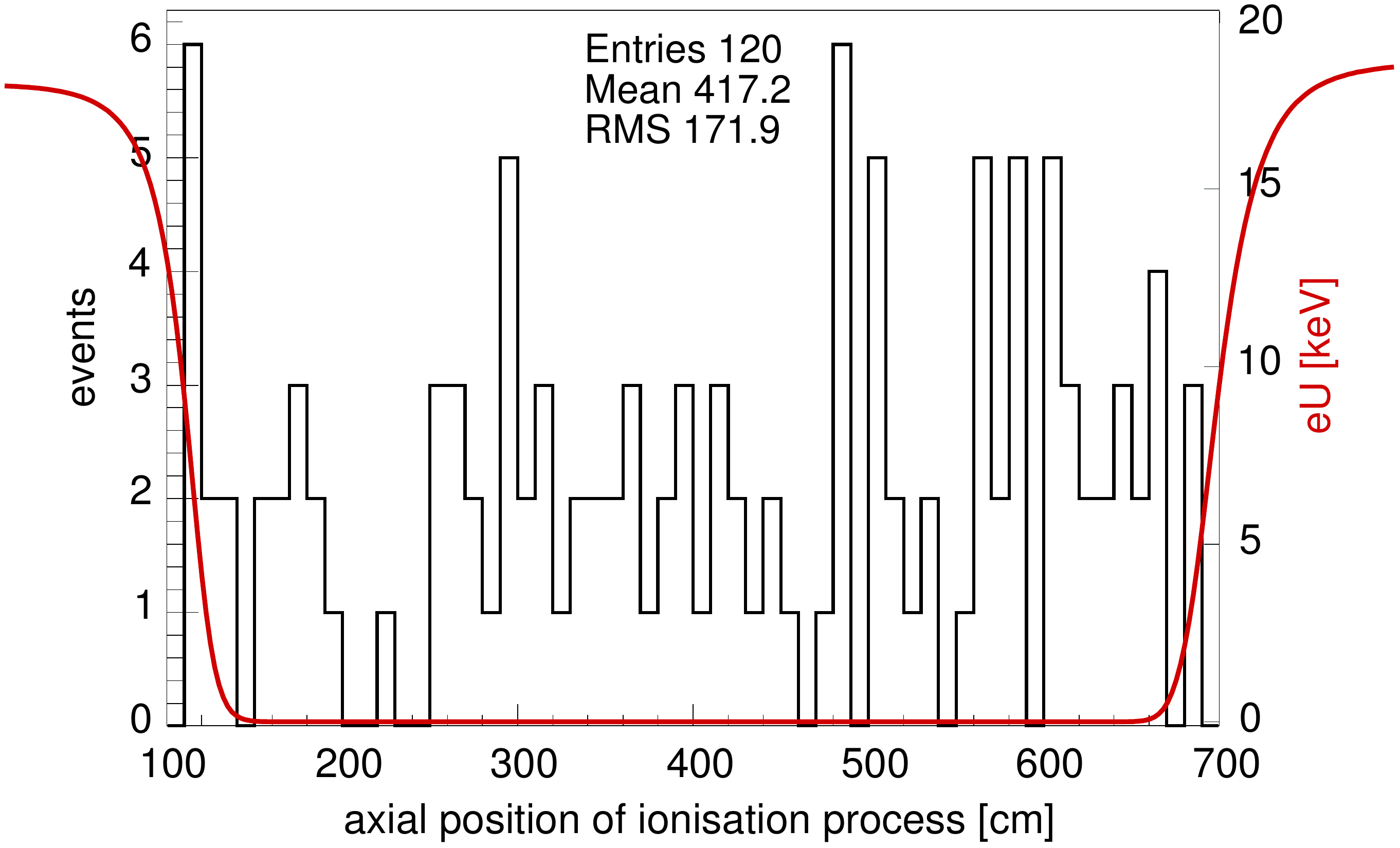}
  \caption{Position of simulated ionisation events between pre- and main spectrometer on the axis \cite{dipl_essig}.
The full line illustrates the corresponding electric potential (right-hand scale).}
  \label{fig:ionization_position}
\end{figure}

Of major concern are the ionisation processes, which occur with a
rate of 1 per 10\,s per stored electron at the considered rest gas
pressure of $10^{-10}$\,mbar. They can start a chain reaction if
the secondary electrons obtain a sufficient amount of energy to allow further
ionisation, in particular through the subsequent acceleration
by the electric field. Figure~\ref{fig:ionization_position} shows the position
on the axis at which the ionisation processes took place for the
10 simulated trapped electrons of
fig.~\ref{fig:ionization_cooling}. Clearly, most
ionisation processes take place at ground potential, where no
potential energy is picked up. However, a small fraction of the
ionisations take place at the end of the trap\footnote{Although the track length at high potential is small, the scattering probability is rather large,  since the electrons are decelerated by the electric potential and have a larger scattering cross section (see fig.~\ref{fig:scattering_xsection}).}. In this
case, the secondary electrons gain energy from the electric
potential and are thus able to further feed the ionisation chain
process. To investigate their influence we would have to track these secondary electrons and their tertiary ionisation products further. 
For the set-up described in the KATRIN design report \cite{kdr} and the
start parameters chosen in the simulations of
figs.~\ref{fig:ionization_cooling} and
\ref{fig:ionization_position}, on average just one secondary electron is
created at a high electric potential per initial trapped
electron. This number becomes smaller for larger starting angles and larger
for smaller starting angles of the initial electron. Therefore, a significant
fraction of initial electrons will create at least one secondary electron and
thus will start a chain reaction with a large number of secondary ions $N_{ion}$.

For the actual KATRIN default design this electron multiplication factor is
significantly larger since the transport magnets have been omitted. Just one
short solenoid with the beam tube on ground potential connects the pre- and
the main spectrometer with their walls on high potential. Consequently the
synchrotron cooling is much less efficient and a lot more secondary electrons
are created by ionisation processes in the regions of high potential at the
end of the trap. In addition, due to the larger track length on high
potential, the probability to generate secondary electrons on high potential
is larger. Still, the above simulations hold qualitatively and a
significant background contribution is expected from this inter-spectrometer
trap at KATRIN.

In order to understand this background contribution quantitatively, a
significantly scaled-up simulation effort is under way. Trapped
electrons and all their secondary and higher scattering products are
tracked using the actual KATRIN design. First results show that many positive
ions are accelerated into the KATRIN main spectrometer, confirming that this
background mechanism is not negligible and requires further investigations and
measures. The details of these more advanced simulations will be discussed
elsewhere \cite{glueck10} after their finalisation. In this work we will now
concentrate on the experimental investigation of this potentially large background.

\begin{figure*}[!hbt]
 \centering
\includegraphics[width=0.9\textwidth]{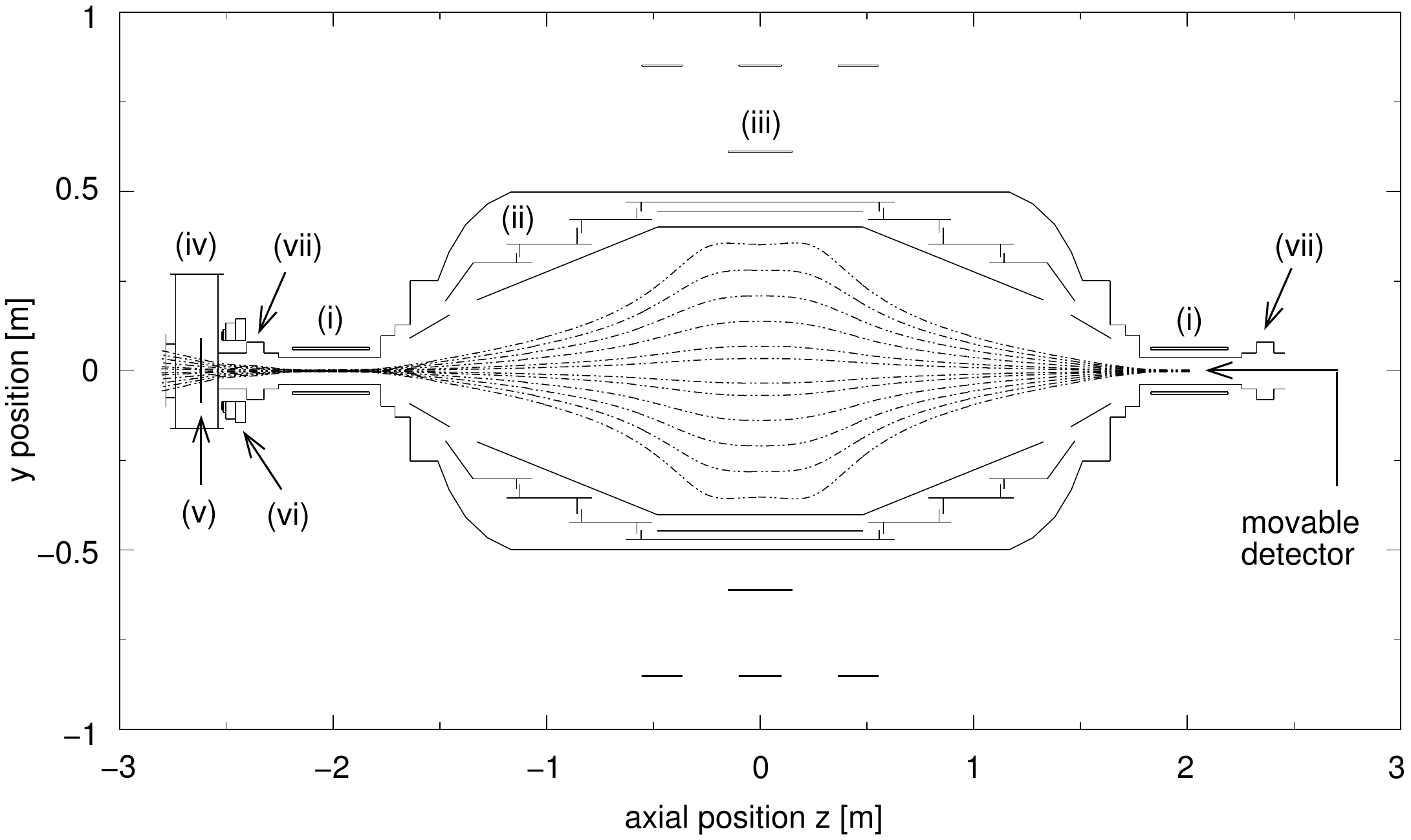}
\caption[Set-up used at the Mainz {\Mac} for testing the trapping conditions
between the KATRIN pre- and main spectrometer]{Set-up used at the Mainz {\Mac}
  for testing the trapping conditions between the KATRIN pre- and main
  spectrometer. (i) superconducting solenoids, (ii) electrode system consisting of  vacuum tank at ground potential, solid and inner wire electrode systems on  high voltage, (iii) field-shaping air coils, (iv) vacuum chamber housing both the sweeping wire installation and (v) the backplate electrode on high voltage, (vi) water-cooled additional coil for local enhancement of the magnetic field at the position of the sweeping wire (not shown in this drawing), (vii) valves. The detector is situated on a movable sleigh close to the solenoid on the right-hand side. Magnetic field lines (settings corresponding to a resolving power of $B_{max}/B_{min}=E/\Delta E \approx 2\cdot 10^4$) are indicated as dashed curves.\label{fig:geometry-mainz}}
\end{figure*}

The most straightforward way to empty the inter-spec\-tro\-me\-ter electron trap
would be to switch off the high voltage temporarily and thus eliminate the
trapping condition for a short time. During data taking this would have to
take place periodically, with short time intervals in between, depending on
the fill rate of the trap. However, the abrupt and large change of
the high voltage from $-18.6~\mathrm{kV}$ to zero and back would be detrimental
to the high stability of the high voltage of the main spectrometer and its
measurement, which both are required to be better than $50~\mathrm{mV}$, \ie,
$3~\mathrm{ppm}$, in order to achieve the required sensitivity for the
neutrino mass at KATRIN of $0.2~\mathrm{eV/c^2}$ \cite{kdr}. Switching the high
voltage of the pre-spec\-tro\-me\-ter down to zero would result in a huge
beta electron flux of $10^{10}~\mathrm{s}^{-1}$ into the main
spectrometer. These electrons would scatter there and induce additional
background. The prevention of this is the main purpose of the KATRIN pre-spectrometer.
Another way to eject trapped electrons would be to use a transversal electric
field, which would remove the electrons by the $\vec E \times  \vec B$-drift
(see eq.~(\ref{eq:drift})). However, this would require impractically high
electric potentials in the case of the KATRIN set-up ($\approx \unit[20]{kV}$, \cite{dipl_essig}).
A third way would be the use of a mechanical wire which is rapidly swept through the trap every few seconds during measurement pauses. During a sweep this wire collects electrons that are created and stored in the trap. A sweep of the wire could be easily achieved by applying an electric current pulse through the wire, which would move the wire by virtue of the Lorentz force in the strong axial magnetic field. In the following, we will report on our investigations of this method.

\section{Experimental set-up}
\label{sec:experimental_setup}

\subsection{Model of the particle trap}

\begin{figure}
 \centering
\includegraphics[height=0.49\textwidth,angle=-90]{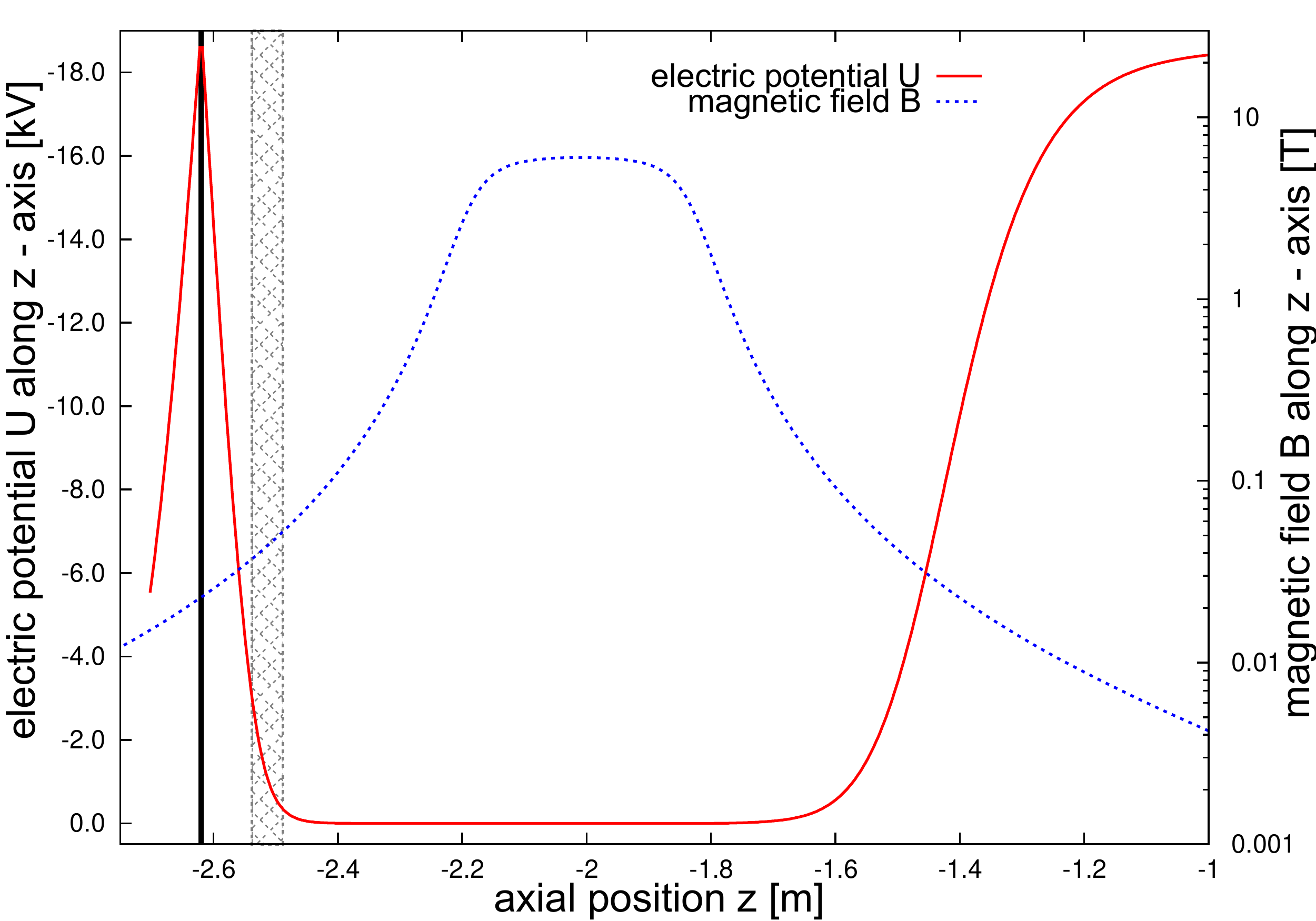}
\caption{Electric potential $U$ (solid line) and magnetic field $B$ (dotted
  line) of the Penning-like trap at the test set-up. the solid vertical bar
  indicates the position of the disc-shaped electrode; the shaded area
  marks the region of the sweeping wire.}
\label{fig:el-magn-fields}
\end{figure}

The pre- and main spectrometer of KATRIN were not yet available for the tests
of this sweeping wire. Therefore, a similar trapping configuration was realised at the spectrometer of the former Mainz neutrino mass experiment
\cite{picard-nimb}, which is also a {\Mac}. The Mainz spectrometer itself was
used as stand-in for the {\KM} (fig.~\ref{fig:geometry-mainz}). The electric
potential of the KATRIN pre-spectrometer was emulated by a disc-shaped,
mechanically polished stainless steel electrode on high voltage in a vacuum
chamber at the entrance of the Mainz \Mac. The magnetic field in between this
electrode and the Mainz spectrometer was mainly defined by the superconducting
entrance solenoid of the spectrometer ($B_\mathrm{max} = \unit[6]{T}$). Thus
the two negative potentials of the backplate electrode and the {\Mac} enclose
a region of more positive potential where a strong magnetic field is present
(fig.~\ref{fig:el-magn-fields}), thus creating a Penning-like trap for
electrons, similar to the case between the pre- and main spectrometer at KATRIN. Typical electron energies at KATRIN are of the order of $\unit[18.6]{keV}$ with spectrometer voltages of $U_\mathrm{spec} \approx \unit[-18.6]{kV}$. For this reason the test measurements with this Penning-like trap were made with potentials in the range $\unit[-15]{kV}$ to  $\unit[-18]{kV}$.

A major difference between the electron trap at the KATRIN pre- and main
spectrometer and the above test set-up are the magnetic field lines which connect the disc-shaped electrode (cathode plate) with the detector in the latter case. At KATRIN, special care was taken to prevent such a situation, since this can lead to an increased background and a feedback mechanism which can lead to discharges.

\subsection{Filling mechanism}

\begin{figure}[!htb]         
 \centering
\includegraphics[width=0.485\textwidth]{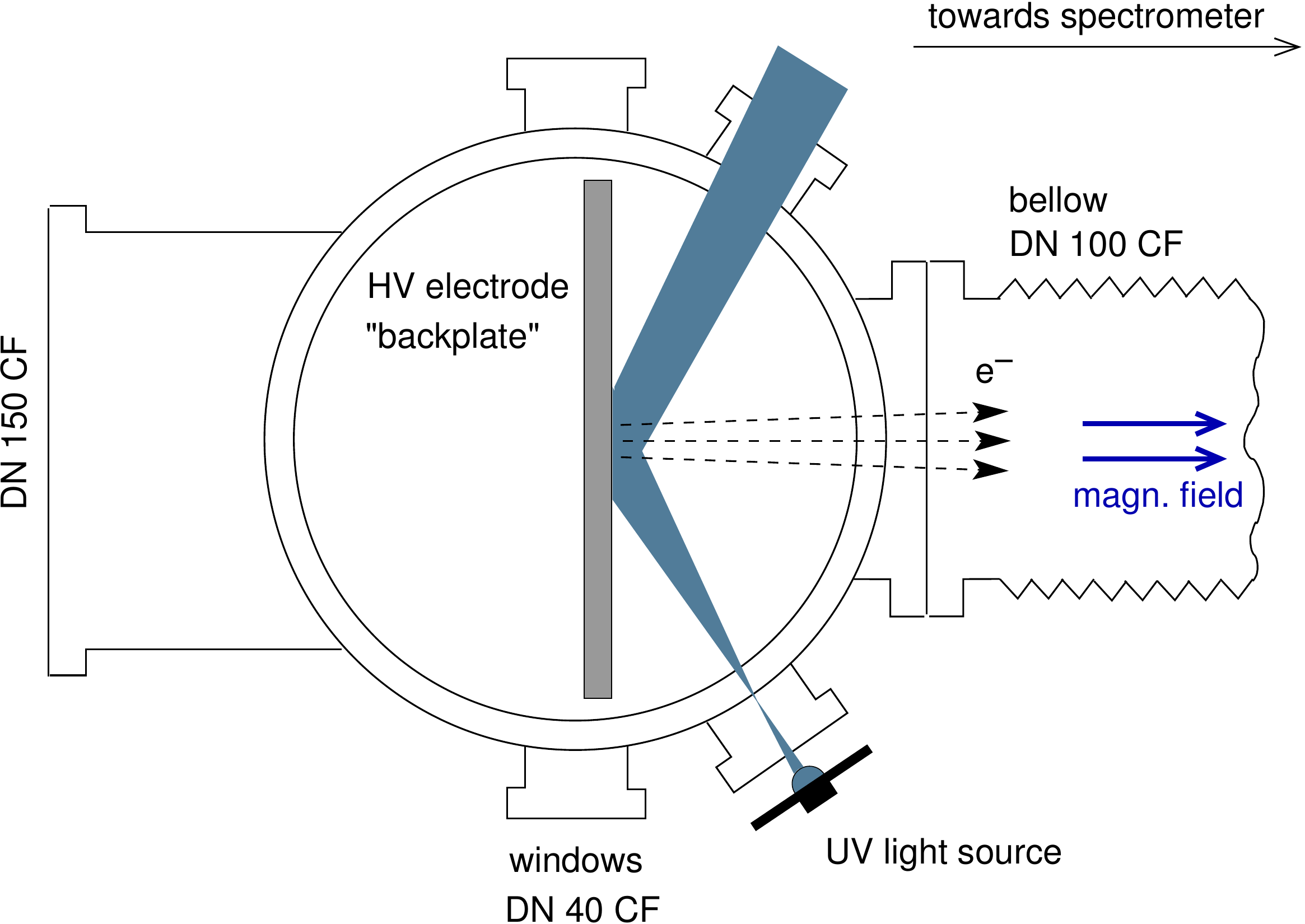}
\caption{Top view of the vacuum chamber housing the disc-shaped high-voltage electrode which serves as a stand-in for the KATRIN pre-spectrometer. Electrons can be created in the centre of this electrode by photoemission using UV-light.
}
\label{fig:ellitopf-topview}
\end{figure}

Another difference of this set-up compared to KATRIN is the lack of electrons
from tritium beta decay\footnote{The tritium source of the Mainz neutrino mass
experiment has been decommissioned some years ago.}, which would provide a filling mechanism for the trap. Two different electron sources were used to fill the trap instead. First, the ''natural'' background of electrons from any
electrode surface on negative high potential, caused by field emission, natural
radioactivity or cosmic ray interactions, can feed the trap. This comprises the
retardation electrodes inside the {\Mac} as well as the disc-shaped
backplate electrode. This mechanism is permanently present and cannot be
switched off. A second filling mechanism was provided by photoelectrons
directly from the backplate electrode (fig.~\ref{fig:ellitopf-topview},
\cite{art_uvled}), which was illuminated by a deep-ultraviolet light emitting
diode (UV-LED, Seoul Semiconductor, types T9B25C and T9B26C
\cite{seoul-led-255,seoul-led-265}, central wavelengths $255~\mathrm{nm}$ and $265~\mathrm{nm}$.). This led to much higher electron numbers
and more efficient and controlled filling of the trap compared to the natural background.

\subsection{The sweeping wire}

\begin{figure}[!htb]
 \centering
\subfigure[sweeping wire used in phase I of the measurements. The wire diameter was $0.4~\mathrm{mm}$.\label{fig:mark1}]{\includegraphics[width=0.485\textwidth]{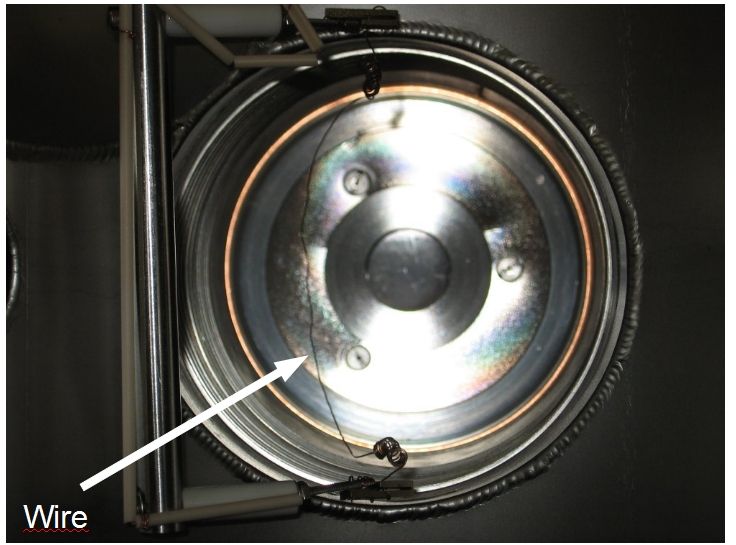}}
\hfill
\subfigure[sweeping wire used in phase II of the measurements. The wire diameter was $1.4~\mathrm{mm}$.\label{fig:mark2}]{\includegraphics[width=0.485\textwidth]{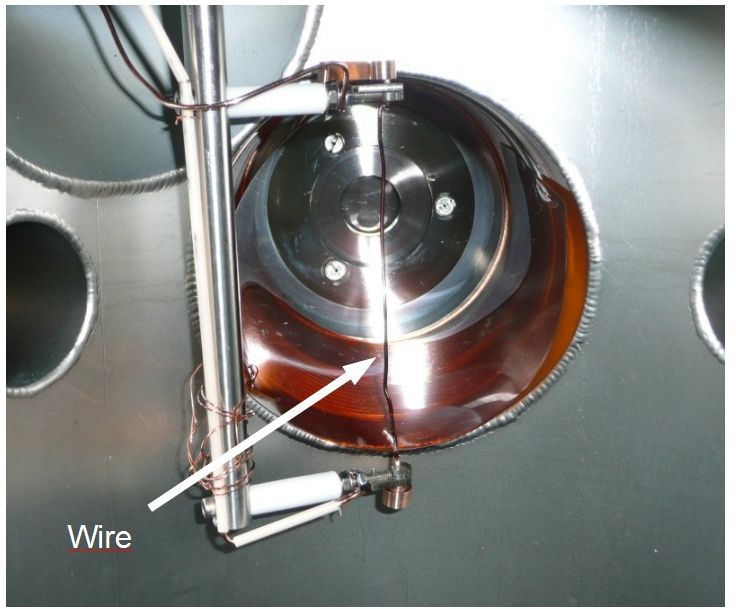}}
\caption[Photographs of two technical realisations of the scanning wire]{Photographs showing two stages of technical realisations of the sweeping wire to empty the Penning-like trap. The pictures were taken from the rear side of the vacu\-um chamber looking from the backplate towards the spectrometer. The line of sight coincides with the central beam axis. In the background of each picture, the bellow connecting vacuum chamber and spectrometer and the closed gate valve towards the spectrometer are visible. As a common feature, both sweeping wire solutions involve springs to facilitate the motion of the wire and provide an electric current via insulated leads, although they differ in the details of mechanical construction. The final construction used flat spiral springs to conduct the electric current and ultra-high vacuum compatible bearings, which enabled stable long-term operations.\label{fig:wirescanner-versions}}
\end{figure}

The basic idea to empty the trap consists of a grounded wire that is periodically moved through the trapping region to collect stored charged particles. It is realised with a semi-circle of a copper wire of $\unit[0.4]{mm}$ or $\unit[1.4]{mm}$ wire diameter, held by suitable supports to allow rotation through the flux tube covering the detector. The wire motion ranges from wall to wall in the enclosing vacuum chamber (DN100). It is moved back and forth through the flux tube by virtue of the Lorentz force due to a periodic current of $\unit[1 - 5]{A}$ through the wire and the local magnetic field of $\unit[0.02 - 0.03]{T}$. In order to facilitate the motion of the sweeping wire and to fit the flux tube, which images the plate onto the detector, inside the vacuum tube, the magnetic field was locally enhanced at the position of the sweeping wire with a water-cooled coil (compare fig.~\ref{fig:geometry-mainz}).
The current driving the wire was provided by
the output of a function generator amplified by a bipolar operational
amplifier (Kepco BOP 20-20M). Using this current the sweeping wire could be
swept swiftly through the trap (rectangular modulation of the current), swept slowly through the
trap (sine wave modulation of the current), placed at the edge of the flux tube (DC current) or
set to the centre of the flux tube (no current). The range of the wire motion  covered the whole
flux tube (wall to wall). In some of the measurements the timing of the sweeping wire was recorded using the trigger output of the function generator. Figure~\ref{fig:wirescanner-versions} shows two sweeping wire configurations in position inside the vacuum chamber, looking from the position of the backplate towards the spectrometer.

\subsection{Spectrometer settings}

The detector used at the exit of the spectrometer to detect electrons from the
plate and the trap was a Si-PIN diode (type Hamamatsu S3590-06) of size $\unit[9
\times 9]{mm^2}$. The magnetic field at the location of the detector was
$B_\mathrm{det}=\unit[0.34]{T}$, corresponding to a covered magnetic flux of $\approx \unit[28]{Tmm^2}$. Since the magnetic field at the plate of the photocathode was $\unit[0.02-0.03]{T}$, an area of $\unit[9-14]{cm}^2$ of the plate was
imaged onto the detector. Typically, measurements were performed at an energy resolution of $\Delta E / E = 1/20000$.

\section{Results of the measurements}
\label{sec:results}

Measurements of the count rate on the detector were performed under various
conditions: without and with trapping conditions present, for the latter case with and without sweeping wire, the sweeping wire in several different operation modes, and for all cases with the spectrometer in transmission as well as closed to low energy electrons from the backplate.

\subsection{Behaviour of the count rate without sweeping wire}

The trap between the backplate and the analysis plane produces several
observable effects, depending on the filling mechanism:

\subsubsection{Without additional filling of the trap}
\label{sec:bgwo}

\begin{figure}[!htb]
 \centering
 \subfigure[without high voltage on backplate]{\includegraphics[width=0.485\textwidth]{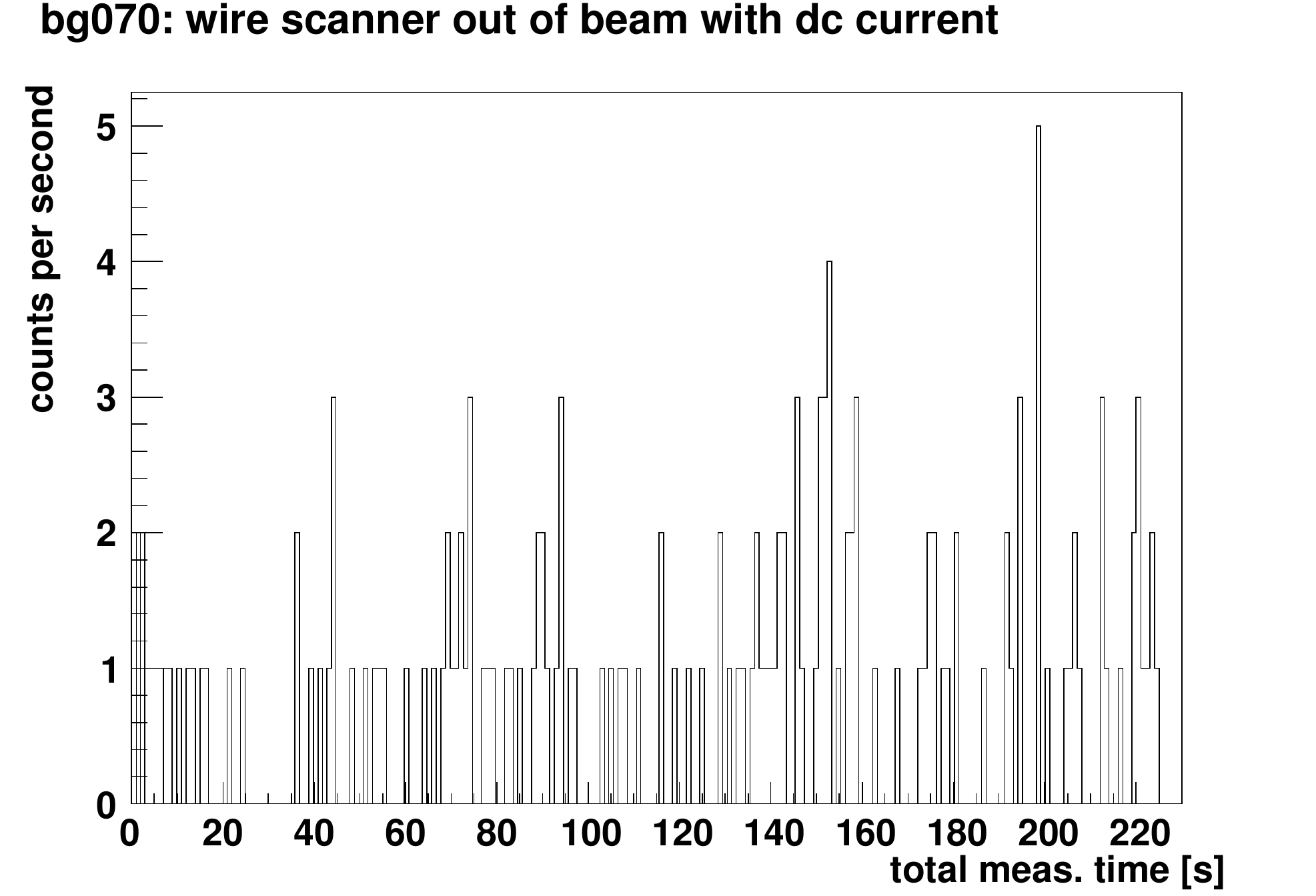}\label{fig:plate-background-a}}
 \hfill
 \subfigure[with high voltage on backplate]{\includegraphics[width=0.485\textwidth]{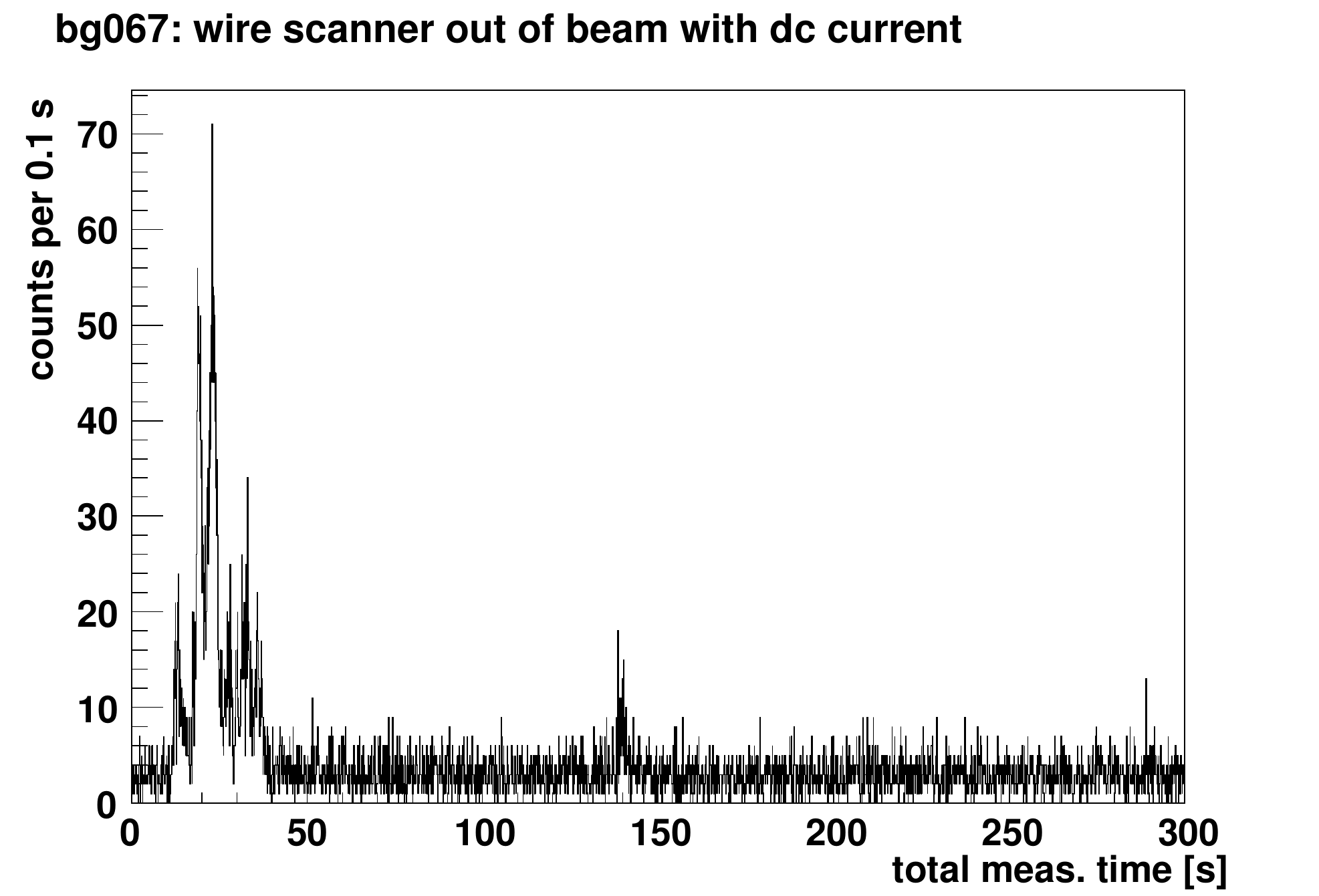}\label{fig:plate-background-b}}
\caption[Count rate without and with high voltage on the backplate]{Evolution of the count rate without and with high voltage on the backplate. In both cases the sweeping wire was positioned outside the magnetic flux tube by applying a constant current; no extra  feeding mechanism of the trap was active.
(a): Count rate for spectrometer background only ($U_\mathrm{plate} = 0$, $U_\mathrm{spec} = -18.6~\mathrm{kV}$). This leads to a count rate of about $\unit[0.7]{\cps}$.
(b): Count rate for $U_\mathrm{plate} = U_\mathrm{spec} + 2~\mathrm{V}$. Strong fluctuations of the count rate on top of a generally elevated level of $25$ to $\unit[35]{\cps}$ are visible (note the difference in the ordinate). \label{fig:plate-background}}
\end{figure}

\begin{figure}[!htb]
 \centering
\includegraphics[width=0.485\textwidth]{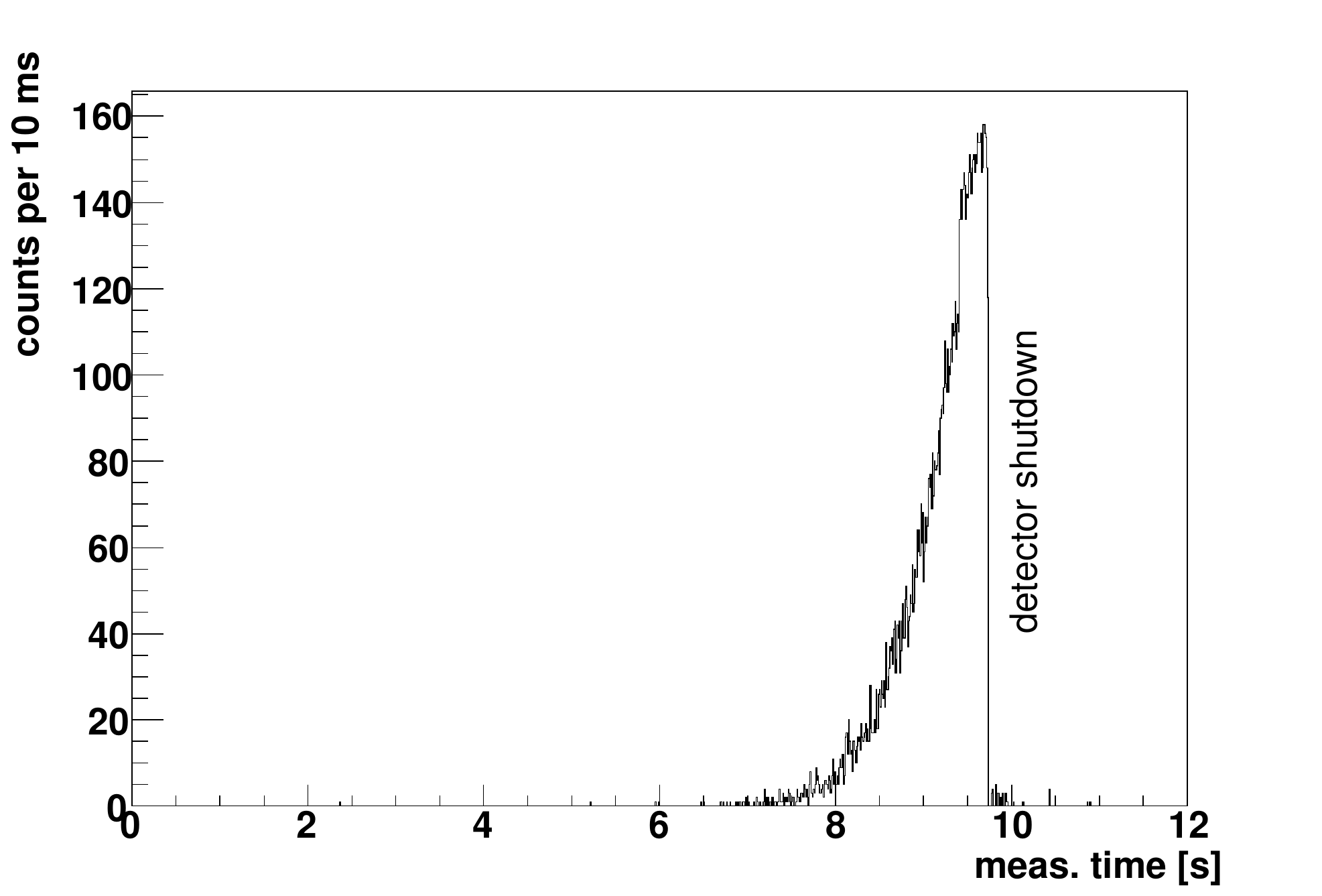}
\caption[Example for the build-up of a discharge and detector shutdown]{Example for the build-up of a discharge for the case of natural background only and without intervention of the sweeping wire. After a quick rise of the count rate to about $\unit[1.6\cdot 10^{4}]{\cps}$ (limited by the data rate accepted by the data acquisition system) the detector was automatically shut down at $t_\mathrm{meas} \approx \unit[9.75]{s}$. \label{fig:sc034-time}}
\end{figure}

Figure~\ref{fig:plate-background} shows the comparison of the count rate for single electrons with and without the trapping condition being present. The trapping condition was switched off by setting the backplate voltage to $U_\mathrm{plate} = \unit[0]{V}$. For the case with the trapping condition being
present the spectrometer voltage was set to a value so that electrons with very low energy from the backplate could not pass the ana\-ly\-sis plane ($U_\mathrm{plate} = U_\mathrm{spec} + 2~\mathrm{V}$ with $U_\mathrm{plate} \approx - \unit[18]{kV}$). In figure~\ref{fig:plate-background-a} the natural background of the spectrometer without the trap is shown. The average count rate in the single electron peak of the detector is $0.7$~\cps. This compares with $25$ to $35$~\cps\ for the case where the trap is present, shown in fig.~\ref{fig:plate-background-b}. In addition to this drastic increase of the count rate frequent and violent bursts were observed when the trap was present. Frequently these led to a strong discharge of the spectrometer and shut down the detector and the spectrometer high voltage due to an excess current (fig.~\ref{fig:sc034-time}).

\subsubsection{With photoelectrons}
\label{sec:pewows}

\begin{figure}[!htb]
\centering
\includegraphics[width=0.525\textwidth]{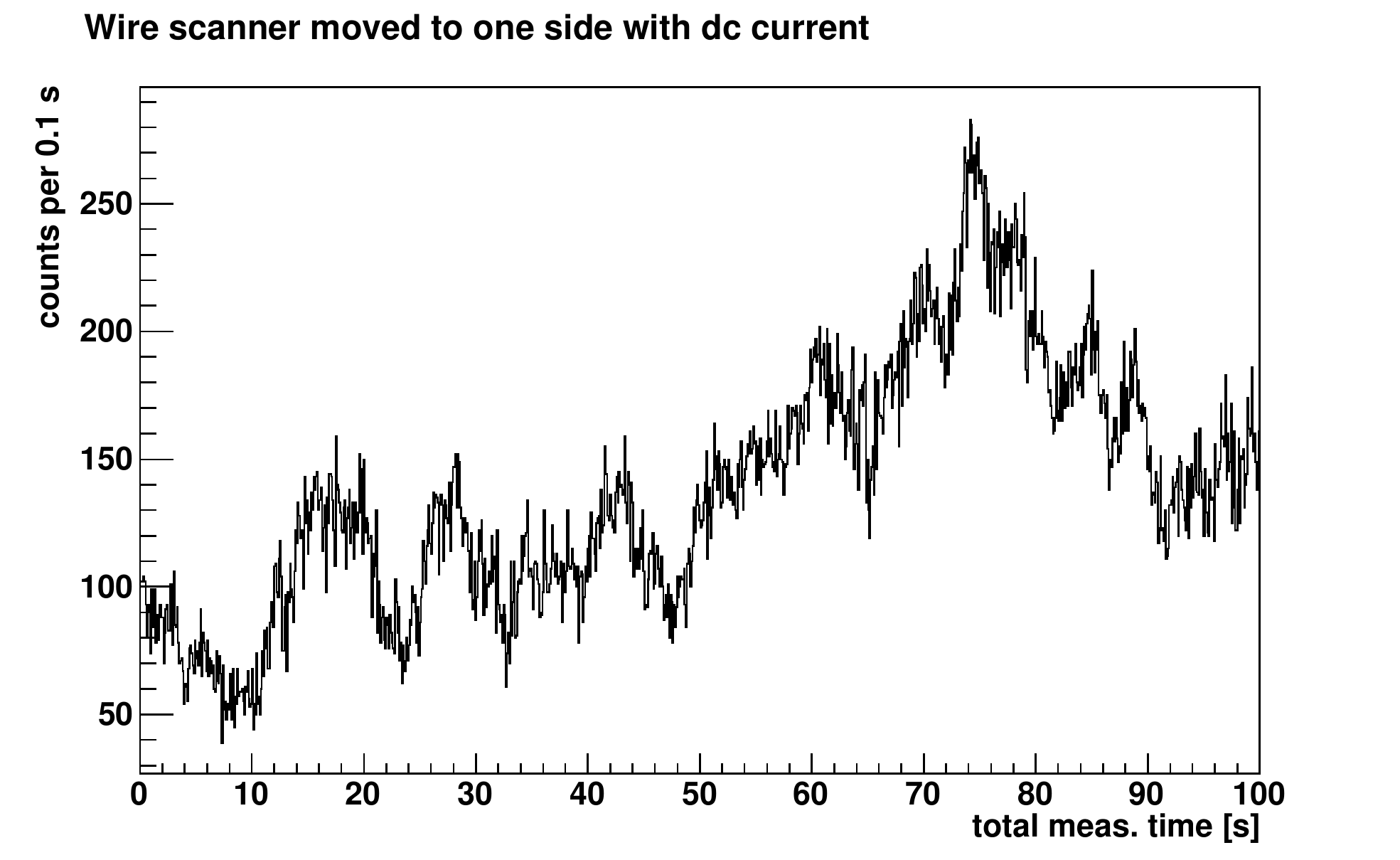}
\caption[Count rate for stopped wire, with UV photoelectrons as filling mechanism]{Count rate with UV  photoelectrons as filling mechanism. Voltage setting: $U_\mathrm{plate} = \unit[-14988.88]{V} \approx  U_\mathrm{spec} + \unit[9.05]{V}$. The sweeping wire was moved out of the magnetic flux tube with a DC current. High count rates of the order of $1\cdot \unit[10^3]{\cps}$ to  $3.5 \cdot \unit[10^3]{\cps}$ were observed. The timing of the events is not correlated with the time of UV emission at the LED.}
\label{fig:scannerfreq-uv-101bins-part3}
\end{figure}

In order to fill the trap in a controlled way photoelectrons were created
with the UV-LED off the backplate. The UV-LED was operated in a pulsed mode
with a pulse duration of $\tau = 12~\mathrm{\mu s}$ and a repetition rate of
1000~Hz. When the spectrometer was operated in transmission the time
structure of the count rate due to this pulsed operation was clearly visible \cite{art_uvled}.
Figure~\ref{fig:scannerfreq-uv-101bins-part3} shows an example of the development of the count rate with the spectrometer
closed ($U_\mathrm{plate} \approx U_\mathrm{spec} + \unit[9.05]{V}$) and the trapping condition being
present. High count rates in the range $1\cdot10^3~\mathrm{\cps}$ to
$3.5\cdot10^3~\mathrm{\cps}$ were observed. Compared to the count rate
due to the natural filling of the trap this is a significant increase of the background. However, no bursts were visible. A potential explanation is that the bursts were hidden in the elevated (and generally variable) background. The timing of the events is not correlated with the pulses of the UV-LED, indicating that the photoelectrons were at least partially trapped and not directly flying to the detector, which is forbidden by energy conservation.

The fill rate of the trap can be estimated from the count rate when the
spectrometer is in transmission.
This yields a fill rate of
$1.1\cdot10^3~\mathrm{electrons/s}$ to $1.4\cdot10^3~\mathrm{electrons/s}$. Since the background during quiet periods with trap and photoelectrons present was
larger by a factor 2-3 than the fill rate of the trap this means that charge
multiplication, a continuous discharge, is taking place in the trap even when
no runaway multiplication with resulting detector shutdown is observed.

In summary, the presence of the Penning-like electron trap between the backplate and the analysis plane led to a significant increase of the background count rate as expected from sect.~\ref{sec:motivation}, to frequent bursts of the count rate and to frequent strong discharges which shut down the spectrometer and the detector.

\subsection{Reduction of background due to the sweeping wire}

The effect of the sweeping wire on the count rate from the trap has been
investigated for rectangular motion of the wire, sine motion of the wire, and
for a wire in fixed position on the axis of the set-up, for the spectrometer electrostatically
closed to direct electrons, and for both filling mechanisms.

\subsubsection{Natural background}
\label{sec:wsnb}

\begin{figure}[!htb]
 \centering
\includegraphics[width=0.485\textwidth]{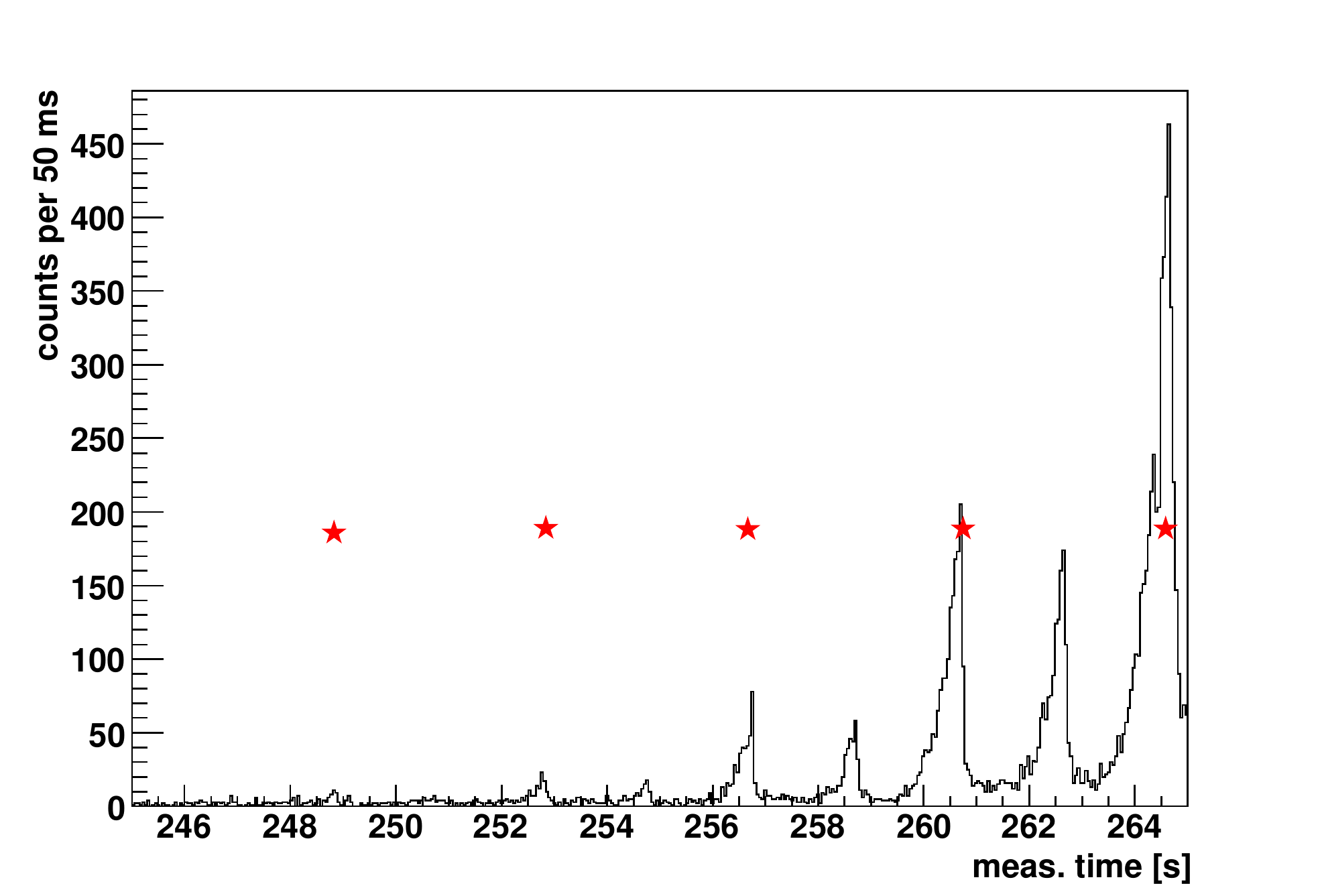}
\caption[Build-up of the count rate at the start of the discharge]{Build-up of the count rate at the start of a discharge for the case of natural background only and the sweeping wire ($1.4~\mathrm{mm}$ wire diameter) active. The trigger signal for the rectangular pulse driving the sweeping wire motion is indicated by red markers. 
The sweep frequency was $f = 0.3~\mathrm{Hz}$. Each trigger signal stands for a sweep of the wire in one direction; the backward sweeps are located in-between two trigger signals at half period and are therefore not shown here. Obviously, the sweeping wire in rectangular mode did not prevent discharges from building up.}
\label{fig:sc033-buildup}
\end{figure}

For natural filling the sweeping wire in rectangular mode reduces the count
rate during its sweep (fig.~\ref{fig:sc033-buildup}), evidence that it removes
charged particles from the trap, but it is not able to
prevent the bursts from building up discharges. This holds even for the
fastest sweeping frequencies that were tested ($2-3~\mathrm{Hz}$). In addition, the fast current pulse is prone to induce electronic noise in the detector.

\begin{figure}[!htb]
\centering
\includegraphics[width=0.485\textwidth]{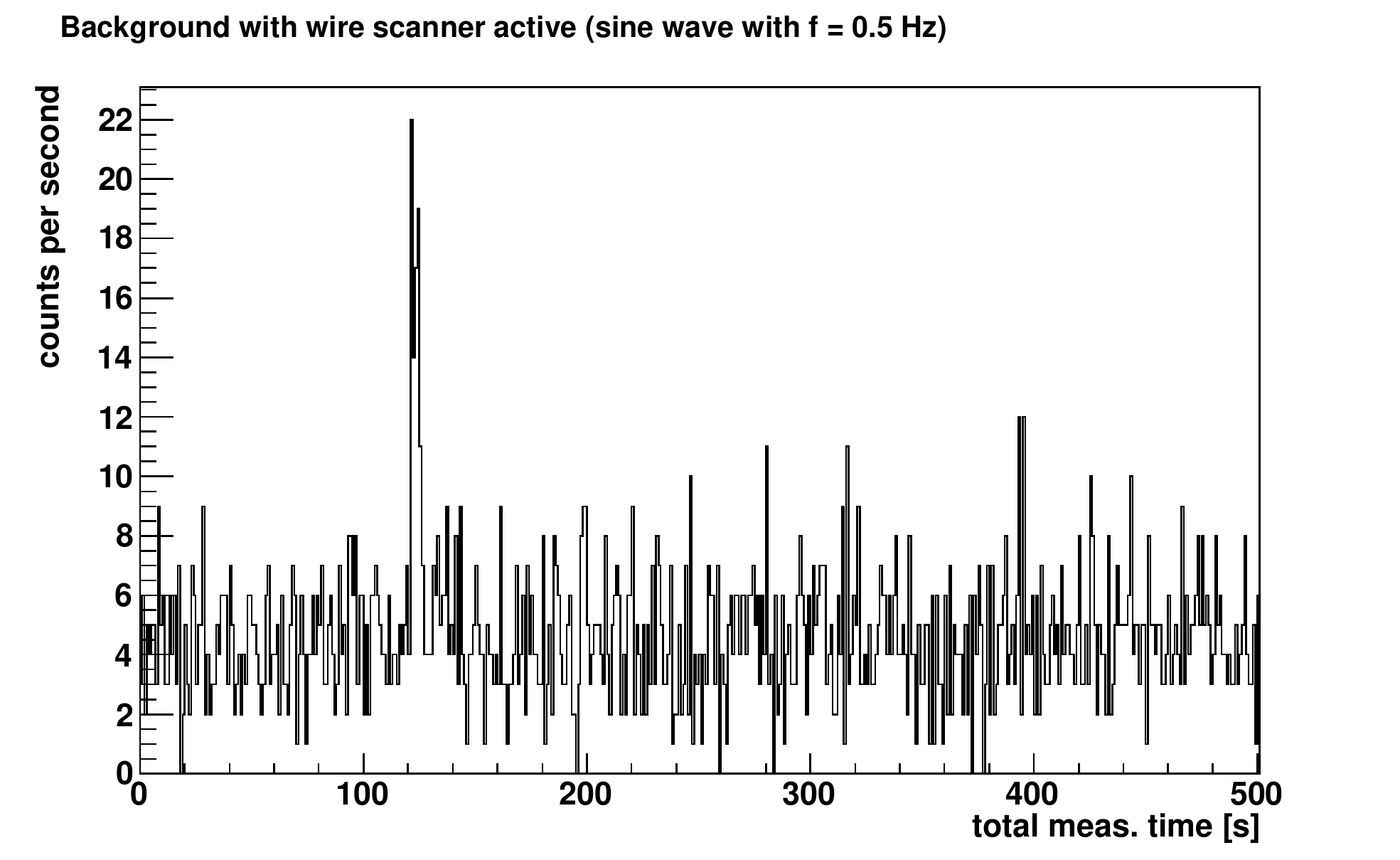}
\caption[Count rate of the sweeping wire]{Count rate for the case of natural background only and the sweeping wire ($0.4~\mathrm{mm}$ wire diameter) in sinusoidal mode with frequency $f = \unit[0.5]{Hz}$.}
\label{fig:scannerfreq-600bins-part2}
\end{figure}

\begin{figure}[!htb]
\centering
\includegraphics[width=0.485\textwidth]{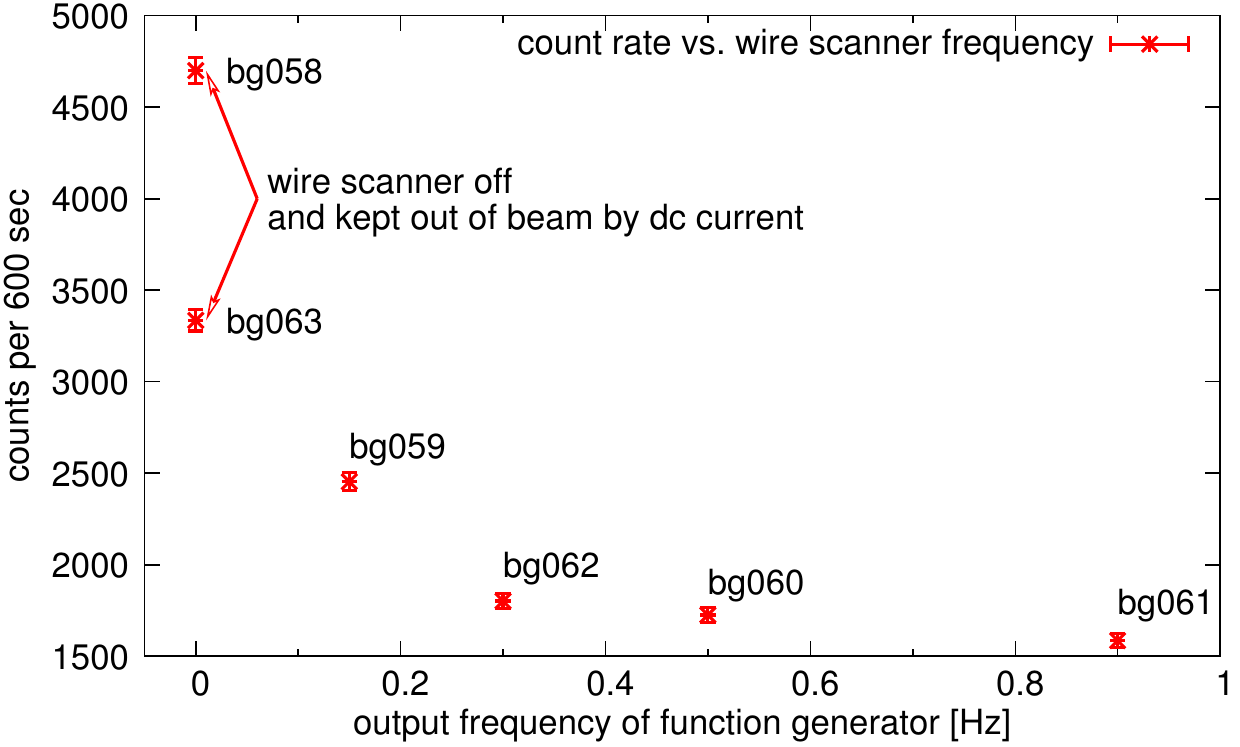}
\caption[Count rate at $15\,\mathrm{kV}$ as a function of the frequency of the sweeping wire motion]{Count rate at $15\;\mathrm{kV}$ for the case of natural background only as a function of the frequency of the sweeping wire ($0.4~\mathrm{mm}$ wire diameter) in sinusoidal mode. The difference in counts observed between the measurements with DC current (bg058 and bg063) can be explained by fluctuations of the bursts in the count rate, see time spectra in fig.~\ref{fig:scannerfreq-600bins-part2}.}
\label{fig:scannerfreq}
\end{figure}

The evolution of the background with time for natural filling
of the trap and a sinusoidal wire motion with frequency
$f=0.5~\mathrm{Hz}$ is shown in fig.~\ref{fig:scannerfreq-600bins-part2}. Both quiet periods and bursts are
still visible at this sweeping frequency. In contrast to the background without sweeping wire (fig.~\ref{fig:plate-background-b}) the overall count rate is reduced and fewer bursts are visible during the measurement time. When the sweep frequency is reduced bursts start to appear more frequently and get stronger. Figure~\ref{fig:scannerfreq} shows the dependence of the total count rate (averaged over quiet periods and bursts) on the sweep frequency. Clearly, the sweeping wire leads to a strong reduction of the count rate, which implies that it removes charged particles from the trap. For the highest sweep frequency no bursts were visible anymore during the time window of the measurement ($\unit[600]{s}$).

Keeping the wire stationary through the centre of the set-up turned out as another effective solution for the suppression of the background and of bursts under the operating conditions used. This method eliminated the strong discharges and most bursts. Only when the spectrometer potential
$U_\mathrm{spec}$ was within several tenths of V of the backplate potential
$U_\mathrm{plate}$ bursts were observed occasionally. The reason why a stationary wire in the centre of the magnetic flux tube works efficiently is the following: Although for this method the wire does not cover the full magnetic flux tube the magnetron motion around the symmetry axis transports the trapped electrons to the wire according to eq.~(\ref{eq:drift}), on time scales much less than a millisecond.

\subsubsection{With photoelectrons}
\label{sec:wswp}

\begin{figure}[!htb]
\centering  
\includegraphics[width=0.485\textwidth]{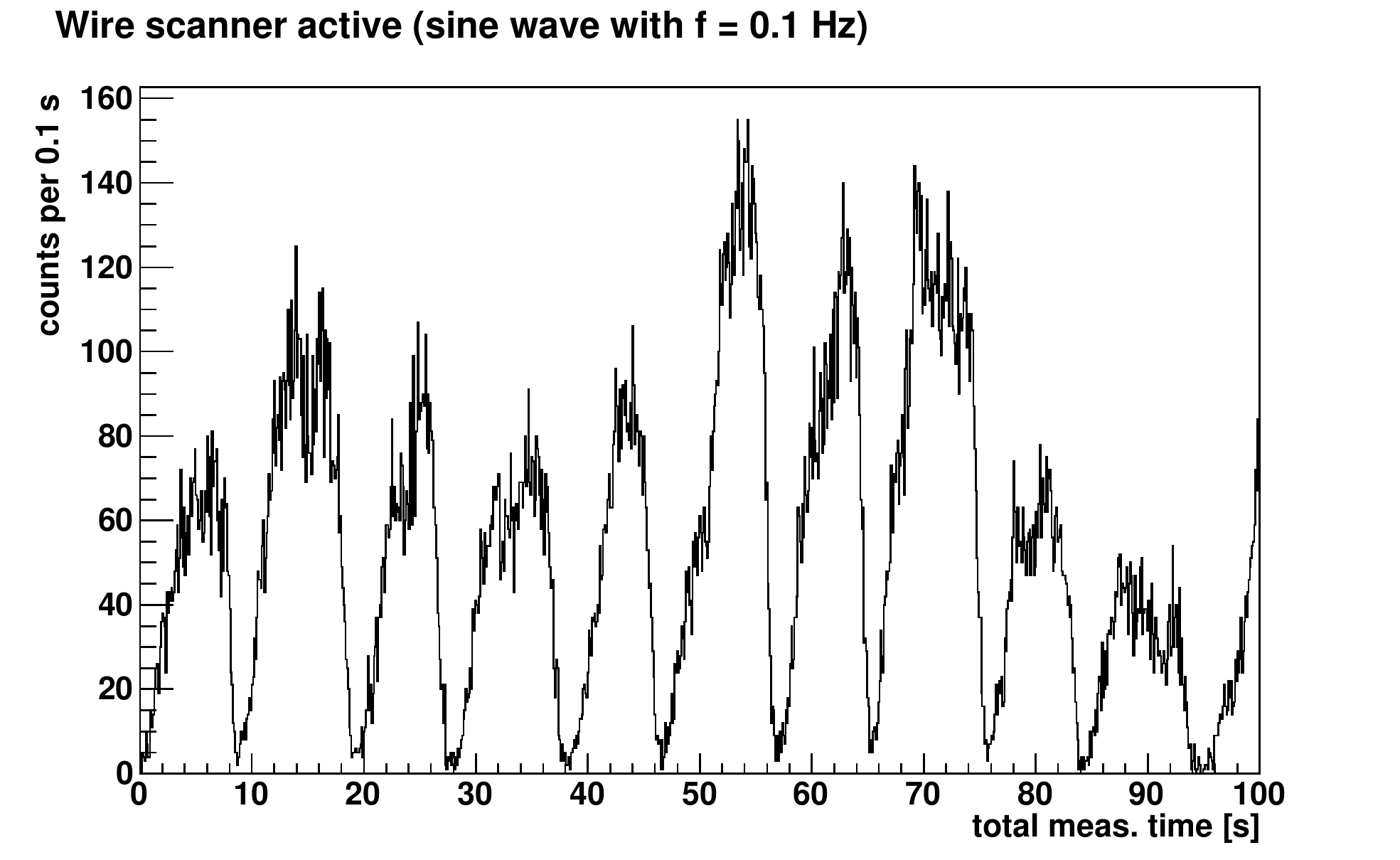}
\caption[Count rate for varying sweeping speeds of the sweeping wire, with UV photoelectrons as filling mechanism]{Count rate with sweeping wire ($0.4~\mathrm{mm}$ wire diameter) in sine mode ($f = 0.1~\mathrm{Hz}$) with UV photoelectrons as filling mechanism. Voltage setting: $U_\mathrm{plate} = \unit[-14988.88]{V} \approx  U_\mathrm{spec} + \unit[9.05]{V}$. The maxima and minima of the count rate correspond to the outer and inner position of the wire, respectively.\label{fig:scannerfreq-uv-101bins-part1}}
\end{figure}

When filling the trap with photoelectrons from the backplate the background with the sweeping wire removed from the trap region was higher by a factor of $\approx 200$ than for the case with natural filling only (compare figs.~\ref{fig:plate-background-b} and \ref{fig:scannerfreq-uv-101bins-part3}).
As in the previous measurements, the sweeping wire in the sine mode results in a significant reduction of the overall rate with the lowest rate being $\approx 100~\mathrm{\cps}$ at the maximum sweep frequency of $0.5~\mathrm{Hz}$. Due to the higher count rate here the sweeping wire motion leads to a direct modulation of the count rate (fig.~\ref{fig:scannerfreq-uv-101bins-part1}). The dependence of the count rate on the frequency of the sweeping motion is also similar, but again at elevated count rates. Having the sweeping wire stationary in the centre reduces the count rate the most to $\approx 10 \mathrm{~counts/s}$.
As a further observation, the arrival times of the electrons were not correlated with the UV-LED pulse timing, in contrast to the case of transmission of the spectrometer \cite{art_uvled}. This confirms that the remaining low count rate stems from secondary ionisation products of electrons which have been stored in the trap.

\subsection{Quenching of discharges}

\begin{figure}[!htb]
 \centering
\includegraphics[width=0.485\textwidth]{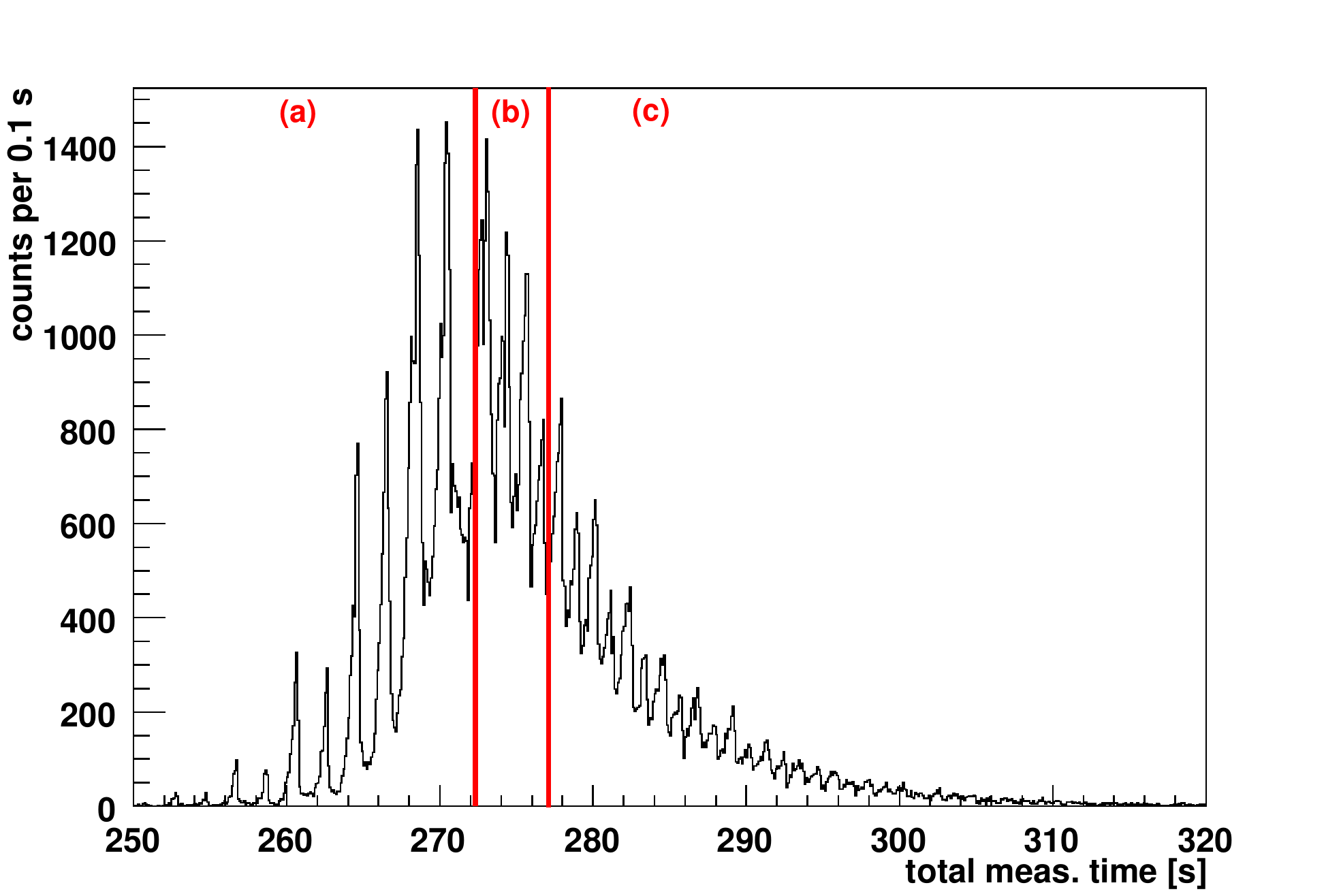}
\caption[Influence of sweeping wire motion on count rate]{Influence of sweeping wire motion on an emerging discharge, with UV-photoelectrons ($1.4~\mathrm{mm}$ wire diameter). The labels (a), (b) and (c) mark intervals with different sweeping wire settings, see table~\ref{tab:sc033-intervals}. Clearly, the sweeping wire in fast sine mode will extinguish discharges.\label{fig:sc033-maindischarge}}
\end{figure}

A major problem caused by the trap were discharges which occur after a variable time (see sects.~\ref{sec:bgwo}, \ref{sec:wsnb}). Figure~\ref{fig:sc033-buildup} shows a typical initial development of such a discharge with the sweeping wire in rectangular mode. After a significant build-up of the discharge the sweeping wire was switched from rectangular mode to slow sine mode at the same frequency (0.3~Hz). This stopped the build-up of the discharge (see fig.~\ref{fig:sc033-maindischarge} and table~\ref{tab:sc033-intervals}) and an additional increase of the frequency (0.5 Hz) fully extinguished it after some time. This behaviour was reproducible and was repeatedly seen for discharges which, under otherwise similar circumstances, would have led to a gigantic discharge without sweeping wire (compare fig.~\ref{fig:sc034-time}), thereby shutting down the detector and the spectrometer high voltage. The rectangular mode by itself was not sufficient to prevent the discharge from increasing.

\begin{table}[!htb]
\centering
\caption[Overview of different measurement intervals in run sc033]{Overview of the sweeping wire operation within the different time intervals for the discharge in in fig.~\ref{fig:sc033-maindischarge}. \label{tab:sc033-intervals}}
\begin{tabular}{|l|l|c|}
\hline
 {\bf meas. time interval}              &       {\bf sweeping wire motion}      \
                         & {\bf label}\\
\hline\hline
0   -- 272 s                            & rectangular, $f \approx \unit[0.3]{Hz\
}$       & (a) \\
272 s -- 277 s                  & sine wave, $f \approx \unit[0.3]{Hz}$        \
 & (b) \\
277 s -- end of run             & sine wave, $f \approx \unit[0.5]{Hz}$        \
 & (c) \\
\hline
\end{tabular}
\end{table}

A very efficient way to quench or prevent discharges from occurring was the
sweeping wire in its centre position. This prevented any strong discharges from
occurring. Small bursts of the count rate were only seen occasionally close to
the edge of transmission.
This option was realised at a later stage with a grounded stationary wire
crossing the flux tube from wall to wall and going through the central
axis. This set up was used to perform test measurements of various
photoelectron sources for the calibration of the KATRIN experiment
\cite{art_uvled,art_fegun,art_asegun}. The efficacy of the stationary wire again shows the importance of the magnetron drift for the stored electrons.

\section{Discussion}
\label{sec:discussion}

It has been shown that a wire, which is swept through a Penning-like trap that creates background electrons by secondary processes outside the trap with a KATRIN-like configuration, reduces that background and can prevent catastrophic discharges caused by this trap. 
The background count rate both with and without (fig.~\ref{fig:scannerfreq}) injecting photoelectrons depends on the sweeping frequency. 
Without injecting photoelectrons the minimum achievable count rate was $3~\mathrm{\cps}$ compared to $\approx 1~\mathrm{\cps}$ without any trap present, 
showing that at low fill rate the sweeping wire is highly efficient.
A rectangular sweeping motion was not sufficient to prevent discharges. This could only be achieved with a sinusodal wire motion. 
Additionally, beginning discharges could even be extinguished by switching to a fast sine sweeping motion.

The effectiveness of the sweeping wire and therefore the achievable background count rate depend on the fill rate and the multiplication factor in the trap, \ie, the ionisation time constant. The fill rate was intentionally high in the case discussed here due to the direct field lines from a cathode through the trap caused by the backplate on negative high voltage inside the flux tube. The multiplication factor depends on the residual gas pressure $p$ and the effective path length of the particles in the trap. The former was several $10^{-9}\;\mathrm{mbar}$ for the measurements discussed here. The effective path length depends on the time scale $\tau_\mathrm{magnetron}$ of the magnetron motion of the electrons in the trap, which determines the time until a stored particle will hit the sweeping wire. For the above measurements this time was determined from simulations to be $\tau_\mathrm{magnetron} \ll 1 \mathrm{s}$. High multiplication factors result from large gas pressure $p$ and large magnetron rotation period $\tau_\mathrm{magnetron}$, low ones are achieved for small $p$ and small $\tau_\mathrm{magnetron}$. For the trap described in figs.~\ref{fig:geometry-mainz} and \ref{fig:el-magn-fields} the sweeping wire was sufficiently effective. It can therefore be expected that for similar and less severe trapping situations (filling mechanism, pressure, magnetron time scale) the sweeping wire will be a useful method to reduce the background due to such a trap and to prevent discharges.

Having the sweeping wire fully stationary in the centre of the trap reduces
the background rate as well. The count rate was $\approx 10 \mathrm{~\cps}$
with a stationary sweeping wire in the centre when the trap is filled by
photoelectrons (sect.~\ref{sec:wswp}). This is a reduction by two orders of
magnitude from the count rate observed for photoelectrons without sweeping
wire (sect.~\ref{sec:pewows}). Comparison of this count rate with the count
rate of $\approx 1~\mathrm{\cps}$ without any trap present
(sect.~\ref{sec:bgwo}) shows that the stationary wire significantly reduces
the background on the detector caused by the electron trap but does not eliminate it fully.
The count rate of $\approx 10~\mathrm{\cps}$ with stationary wire and with the injection of photoelectrons compares at similar conditions 
with $\approx 100~\mathrm{\cps}$ with the sweeping wire at the highest sweep frequencies and with $\approx 30~\mathrm{\cps}$ 
without both photoelectrons and without any sweeping wire (sect.~\ref{sec:bgwo}), again showing the high effectiveness of the stationary wire. 
As a consequence of these results a stationary sweeping wire was used for all further experiments at the spectrometer of the former Mainz neutrino mass experiment when the electron trap was present
and when the stationary wire in the centre of the beam tube does not harm the measurements. 
This reliably prevented discharges from occurring for fill rates comparable to the ones discussed here and led to sufficiently low count rates for 
these experiments (see, \eg, \cite{diss_kathrin,art_uvled}).

\subsection{Application in the \KE}

In contrast to the disc-shaped electrode used here, the inter-spectrometer trap at the \KE \ has no cathode connected via magnetic field lines to feed it or to provide an additional amplification mechanism. It will be filled by electrons created in the surrounding electrodes by cosmic rays and natural radioactivity, as well as by electrons from tritium beta decay, which may scatter on residual gas atoms and thus get trapped. The residual gas pressure will be significantly lower than in the above measurement, at $p_\mathrm{KATRIN}=10^{-11}\mathrm{mbar}$,  and the magnetron motion is 
calculated to be in the range of $\tau_\mathrm{magnetron} = 20 - 200~\mu \mathrm{s}$, comparable with the magnetron motion for the Mainz set-up. This could lead to more favourable multiplication factors at KATRIN than at the above measurements and the efficacy of the sweeping wire could even be better. However, this will strongly depend on the filling mechanism of the trap and must eventually be determined experimentally. 

The sweeping wire could be used both in sweeping mode as well as a stationary wire. A stationary wire would form a permanent obstacle and lead to the scattering of electrons\footnote{It would be a mechanical obstacle: since it is located in between the spectrometers, where the electric potential is close to zero, and the wire itself is on ground potential it should not influence the trajectories of electrons from tritium decay that do not hit it directly. However, a fraction of $\approx 17\%$ of the pixels of the KATRIN Si-PIN detector would be covered by the shadow of the wire.}. Scattered electrons cannot be used for the measurement of the tritium energy spectrum, meaning that the detector pixels on which the sweeping wire is imaged have to be discarded, leading to a loss of statistics. A sweeping wire operated in sine mode would have the same problem, but has the advantage that it could be fully removed from the beam or centred in the flux tube like a stationary wire. However, if the electron multiplication in the trap should be sufficiently small then the sweeping wire could be operated with pauses between two periods of the sine or even in rectangular mode. The measurement of the tritium beta spectrum would take place while the sweeping wire is outside the flux tube imaged onto the detector. Similarly a stationary wire could be inserted into the centre of the beam tube periodically
separated by time intervals without wire.
Again, this has to be investigated experimentally in the final configuration
at the KATRIN set-up.

\subsection{Outlook: potential applications at other experiments}

Besides the KATRIN experiment there are other experiments which suffer from unintended particle traps causing an increase of the background level or discharges (\eg, in the area of fundamental interaction investigations the WITCH experiment \cite{Bec03,Koz08}, the aSPECT experiment \cite{Glu05} and the NAB experiment \cite{Poc09}). The sweeping wire technique discussed here may also be of use at these experiments.

\begin{acknowledgement}
This work was supported by the German Federal Ministry of Education
and Research under grant number 05 CK5 MA/0. One of the authors (M.\,Z.) was also supported by the MSMT, Czech Republic under the contract LA318. We wish to thank the members of AG Quantum/Institut f\"ur Physik, Johannes Guten\-berg-Universit\"at Mainz, for their kind hospitality and for giving us the opportunity to carry out these measurements in their laboratory. 
\end{acknowledgement}
%


%
\end{document}